\documentclass[aps,pre,twocolumn]{revtex4} 
\usepackage{hyperref}
\usepackage{graphicx}
\usepackage{amsmath}
\usepackage{dcolumn}
\usepackage[utf8]{inputenc}
\usepackage{amssymb}

\begin{document}

\title{3D MHD simulation of linearly polarised Alfven wave dynamics in 
Arnold-Beltrami-Childress magnetic field}

\author{D. Tsiklauri}
\affiliation{School of Physics and Astronomy, Queen Mary University of London, London, E1 4NS, United Kingdom}

\begin{abstract} 
Previous studies (e.g. Malara et al ApJ, 533, 523 (2000)) considered small-amplitude 
Alfven wave (AW) packets in Arnold-Beltrami-Childress (ABC) magnetic field using
WKB approximation. They draw a distinction between 
2D AW dissipation via phase mixing and 3D AW dissipation via exponentially divergent 
magnetic field lines. In the former case AW dissipation time scales as $S^{1/3}$ and 
in the latter as $\log(S)$, where $S$ is the Lundquist number. In this work  
linearly polarised Alfven wave dynamics in ABC magnetic field
via direct 3D MHD numerical simulation is studied for the first time.
A Gaussian AW pulse with length-scale much shorter than ABC domain length and a harmonic
AW with wavelength equal to ABC domain length are studied for four different resistivities.
While it is found that AWs dissipate quickly in the ABC field, 
contrary to an expectation, it is found the
AW perturbation energy increases in time. 
In the case of the harmonic AW the perturbation energy growth is transient in time, 
attaining peaks in both velocity and magnetic
perturbation energies within timescales much smaller than the resistive time.
In the case of the Gaussian AW pulse the velocity perturbation energy 
growth is still transient in time, 
attaining a peak within few resistive times, while magnetic perturbation energy
continues to grow.
It is also shown that the total magnetic energy decreases 
in time and this is governed by the resistive evolution 
of the background ABC magnetic field rather than AW damping. 
On contrary, when the background magnetic field is uniform, 
the total magnetic energy decrease is prescribed by AW damping, 
because there is no resistive evolution of the background.
By considering runs with different amplitudes and by 
analysing  the perturbation spectra, 
possible dynamo action by AW perturbation-induced peristaltic flow and inverse cascade of
magnetic energy have been excluded. Therefore, the perturbation energy growth 
is attributed to a new instability.
The growth rate appears to be dependent on the value of the resistivity 
and the spatial scale of the AW disturbance.
Thus, when going beyond WKB approximation,  AW damping, 
described by full MHD equations, does not guarantee
decrease of perturbation energy. This has implications for the 
MHD wave plasma heating in
exponentially divergent magnetic fields.
\end{abstract}

\maketitle

\section{Introduction} 
Damping of magnetohydrodynamic (MHD) waves is of importance to the solar coronal 
heating problem (e.g. Ref.\cite{2005psci.book.....A} and references therein)
and Tokamak plasmas \cite{2013PhPl...20i2107H,2013PhPl...20h2502P,2013PhPl...20h2132F}.
Phase mixing of harmonic Alfven waves (AW), which propagate in  plasma which has a 
density inhomogeneity in  transverse to the uniform background magnetic field (UBMF) direction,
results in their fast damping in the density gradient regions. 
In this case the dissipation time scales as 
$\tau_D \propto S^{1/3}$. Where $S= L V/\eta \propto 1/\eta$ is the Lundquist number. 
$\eta=1/(\mu_0 \sigma)$ is plasma resistivity, while
$L$ and $V$ are characteristic length- and velocity- scales of the system. 
This is a consequence of the fact that AW amplitude damps in time as
$B_y(x,z,t)\propto \exp(-\eta C^\prime_A(x)^2 t^3 k^2/6)$, 
where symbols have their usual meaning and $C^\prime_A(x)$ denotes Alfven speed derivative 
in the density inhomogeneity direction
\cite{1983A&A...117..220H}.
Phase mixing of Alfven waves which have Gaussian profile along the background 
magnetic field results in somewhat slower, power-law damping,
$B_y\propto t^{-3/2}$, as established in
Ref.\cite{2002RSPSA.458.2307W}, whilst more elegantly (in mathematical sense) derived in 
Ref. \cite{2003A&A...400.1051T}.
In a different physical contexts it was shown that exponentially diverging magnetic field lines
provide even faster damping 
$B_y=\exp\left(-A_1\exp(A_2 t)\right)$ \cite{1989ApJ...336..442S,2000A&A...354..334D,2007A&A...475.1111S},
resulting  in the wave damping timescale as $\tau_D \propto \log(S)$.

Ref.\cite{2000ApJ...533..523M} considered small-amplitude 
AW packets in WKB approximation in Arnold-Beltrami-Childress (ABC) magnetic field, which
for certain set of parameters and in known regions of 
space possesses property of exponentially diverging magnetic fields. 
Using WKB reduced version of MHD equations they have convincingly demonstrated that 
when a random number of AW packets are injected in the said ABC field, two distinct populations emerge:
(i) ones that dissipate quickly whose damping time  $\tau_D \propto \log(S)$ and (ii) slowly dissipating
AW packets whose damping time scales as $\tau_D \propto S^{1/3}$. Moreover, they have established that 
quickly dissipating AW packets can be associated with damping in exponentially divergent magnetic field
regions of the simulation domain, while slowly dissipating AW packets damp in 
smooth, magnetic-flux-tube-like regions of space. Also, exponentially diverging magnetic fields
were discussed in the context of magnetic reconnection \cite{2012PhPl...19k2901B,2012PhPl...19i2902B}.

To our knowledge the present study is the first that 
investigates AW damping in ABC magnetic fields using a general,
rather than WKB version of 3D resistive MHD equations.
Therefore the present work can account for (i) time evolution of the background magnetic field
and (ii) the effect of launched Alfvenic waves on the physical system.
The latter is not at all a trivial matter, as there are works \cite{2008A&A...490..353G} 
that show in the solar coronal plasma context that
directly coupling the low beta coronal evolution to prescribed 
photospheric motions of the magnetic footpoints 
allows strong magnetic energy accumulation in the corona. They argue that
this amounts to ignoring a possible feedback from coronal loops on photospheric motions. 
However, the energy injected into the corona comes from the photosphere, so in principle the 
coronal loop might act as a conduit communicating photospheric dynamics from one region to another. 

Section II presents the model and results. Section III summarises the main findings.

\section{The model and results}

The numerical simulations presented here are performed using Lare3d \cite{2001JCoPh.171..151A} -- 
a Lagrangian remap code for solving non-linear MHD equations in 3D spatial dimensions. 
The code is second order accurate 
in space and time. The use of shock viscosity and gradient limiters make
the code ideally suited to shock calculations. The code is available for download from 
\url{http://ccpforge.cse.rl.ac.uk/gf/project/lare3d/}.

The considered numerical runs with their identifying names used throughout this paper are shown in
Table~\ref{runs}.
\begin{table}
\caption[]{Numerical simulation summary table:
Columns from left to right indicate:
(i) numerical simulation run identification,
(ii) type of background magnetic field used,
(iii) AW perturbation type ("pulse" stands for "Gaussian pulse"
and "harm." stands for "harmonic"),
(iv) resistivity $\hat \eta$ (in units of $\mu_0 L_0 C_A$),
(v) end simulation time (in units of $\tau_A$),
(vi) pertaining movie numbers,
(vii) pertaining figure numbers.
}
\label{runs}
$$ 
\begin{array}{lllllll}
\hline
\noalign{\smallskip}
\mathrm{Run} &  \mathrm{Backgr.} & \mathrm{Perturb.} & \mathrm{Resis-} &  t_{end} & \mathrm{Movies} &\mathrm{Figs.}\\
\mathrm{ID}  &  \mathrm{field} & \mathrm{type} & \mathrm{tivity} &  [\tau_A] & \mathrm{} &\mathrm{}\\
\noalign{\smallskip}
\hline
\mathrm{con_e} & \mathrm{const}||z & \mathrm{none} & 5\times 10^{-4} & 20 & 1,2 & 1,2\\
\mathrm{con_p} & \mathrm{const}||z & \mathrm{pulse} & 5\times 10^{-4} & 20 & 1 & 1\\
\mathrm{con_h} & \mathrm{const}||z & \mathrm{harm.} & 5\times 10^{-4} & 20 & 2 & 2\\
\mathrm{abc_e0} & \mathrm{ABC} & \mathrm{none} & 0 & 20 & 1 & 3\\
\mathrm{abc_e1} & \mathrm{ABC} & \mathrm{none} & 5\times 10^{-4} & 10\pi & 3,4,5 & 3,4,6,7,9,10\\
\mathrm{abc_e2} & \mathrm{ABC} & \mathrm{none} & 1\times 10^{-4} & 10\pi & \mathrm{none} & 6,7\\
\mathrm{abc_e3} & \mathrm{ABC} & \mathrm{none} & 5\times 10^{-5} & 10\pi & \mathrm{none} & 6,7\\
\mathrm{abc_e4} & \mathrm{ABC} & \mathrm{none} & 1\times 10^{-5} & 10\pi & \mathrm{none} & 6,7\\
\mathrm{abc_p1} & \mathrm{ABC} & \mathrm{pulse} & 5\times 10^{-4} & 10\pi & 4 & 6,8,9\\
\mathrm{abc_p2} & \mathrm{ABC} & \mathrm{pulse} & 1\times 10^{-4} & 10\pi & \mathrm{none} & 6\\
\mathrm{abc_p3} & \mathrm{ABC} & \mathrm{pulse} & 5\times 10^{-5} & 10\pi & \mathrm{none} & 6\\
\mathrm{abc_p4} & \mathrm{ABC} & \mathrm{pulse} & 1\times 10^{-5} & 10\pi & \mathrm{none} & 6\\
\mathrm{abc_h1} & \mathrm{ABC} & \mathrm{harm.} & 5\times 10^{-4} & 10\pi & 5 & 7,10\\
\mathrm{abc_h2} & \mathrm{ABC} & \mathrm{harm.} & 1\times 10^{-4} & 10\pi & \mathrm{none} & 7\\
\mathrm{abc_h3} & \mathrm{ABC} & \mathrm{harm.} & 5\times 10^{-5} & 10\pi & \mathrm{none} & 7\\
\mathrm{abc_h4} & \mathrm{ABC} & \mathrm{harm.} & 1\times 10^{-5} & 10\pi & \mathrm{none} & 7\\
\mathrm{abc_a1} & \mathrm{ABC} & \mathrm{pulse} & 5\times 10^{-4} & 10\pi & \mathrm{none} & 8\\
\mathrm{abc_a2} & \mathrm{ABC} & \mathrm{pulse} & 5\times 10^{-4} & 10\pi & \mathrm{none} & 8\\
\noalign{\smallskip}
\hline
\end{array}
$$ 
\end{table}

\begin{figure}[htbp] 
\begin{center}
\includegraphics[width=8.5cm]{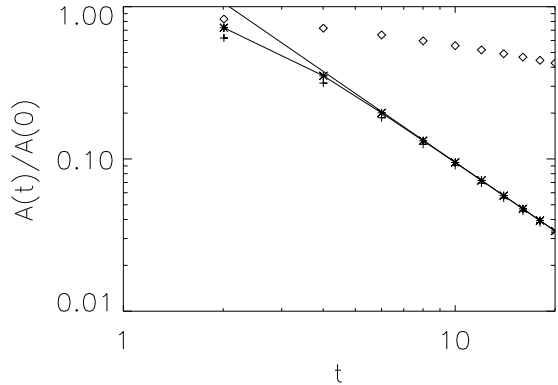}
\end{center}
\caption{Time evolution of an amplitude, normalised to its initial value, for the case of Alfven 
pulse in uniform magnetic field. 
The thin solid line corresponds to the asymptotic solution for large times, Eq.(1). A more general analytical form 
Eq.(2) is plotted with stars connected by thick line. Crosses and open diamonds are numerical simulation results in the
strongest density gradient point $x=(155/512)\times (2\pi)=1.90214$ and 
away from the gradient $x=(1/512)\times (2\pi)=0.0122718$ (first grid cell in $x$-direction), respectively.}
\label{fig1}
\end{figure}

\begin{figure}[htbp] 
\begin{center}
\includegraphics[width=8.5cm]{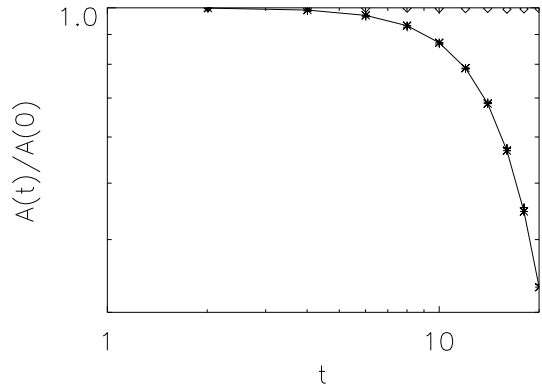}
\end{center}
\caption{As in Fig.(1) but for the case of harmonic AW, except for there is no thin solid line and 
an analytical solution is according to  
Eq.(3) (stars connected by thick line).
The crosses (simulation data) practically overlap with the analytical solution.}
\label{fig2}
\end{figure}

\begin{figure}[htbp] 
\begin{center}
\includegraphics[width=8.5cm]{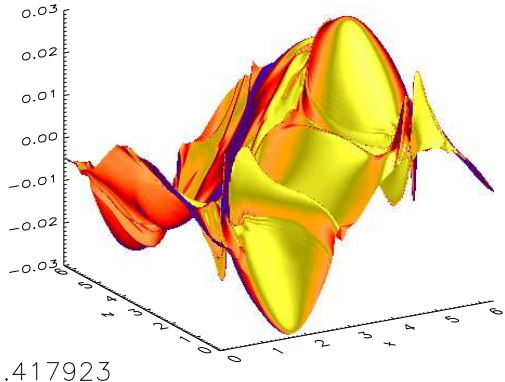}
\end{center}
\caption{$B_y(x,y=y_{max}/2,z,t_{end})-B_{y0}(x,y=y_{max}/2,z,t_{end})$ shaded surface plot,
i.e. difference between the magnetic field y-component in the case of ABC field without  AW pulse but with
resistivity $\hat \eta=5\times 10^{-4}$, i.e. numerical run $\mathrm{abc_e1}$, 
(denoted by $B_y(x,y=y_{max}/2,z)$ in the animated version of this figure i.e. Movie 3 from Ref.\cite{mov}) 
and  magnetic field y-component in the case of ABC field without  AW pulse and without
resistivity $\hat \eta=0$, i.e. numerical run $\mathrm{abc_e0}$, (denoted by $B_{y0}(x,y=y_{max}/2,z)$ in Movie 3).}
\label{fig3}
\end{figure}

\begin{figure*}[htbp] 
\begin{center}
\includegraphics[width=5.5cm]{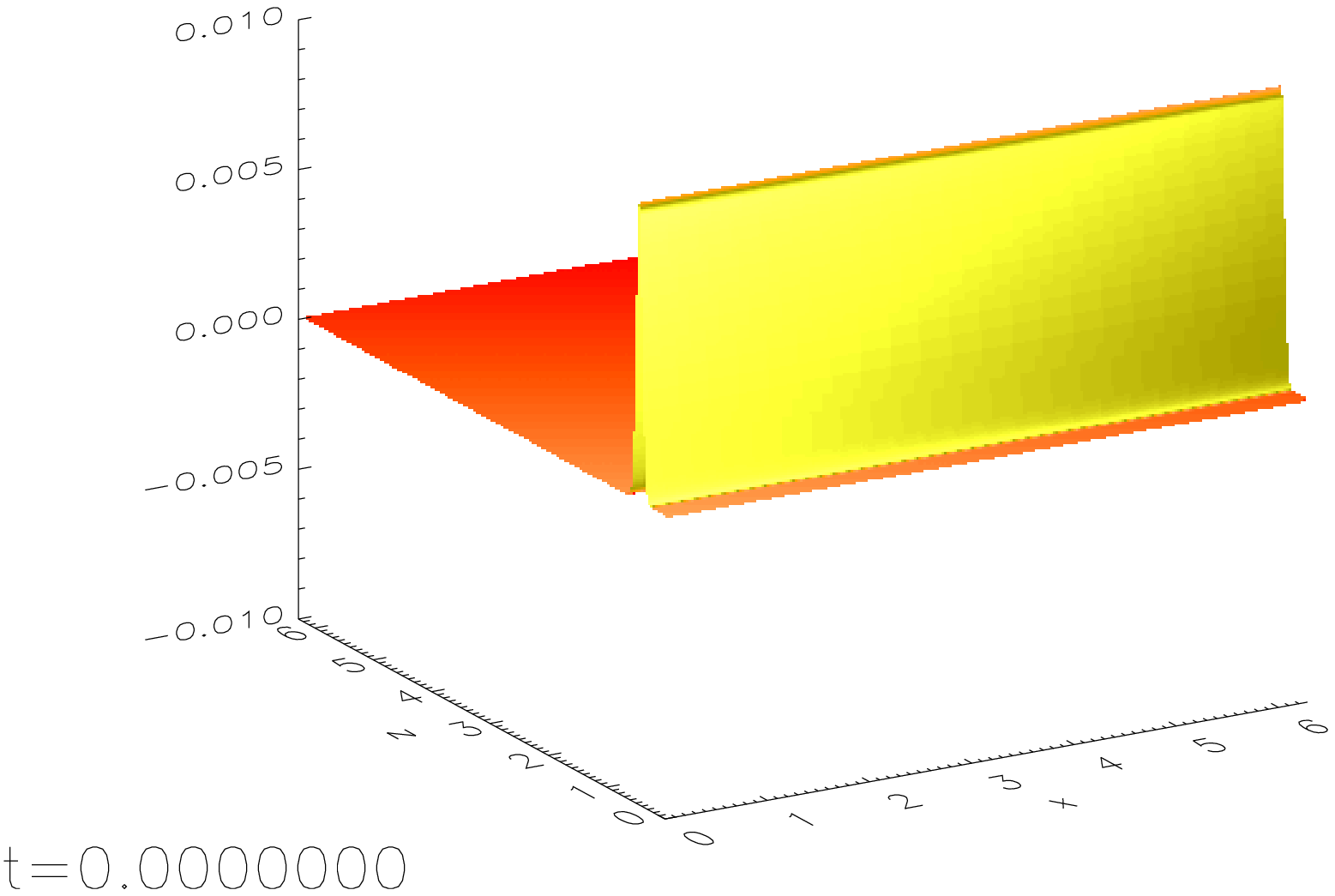}
\includegraphics[width=5.5cm]{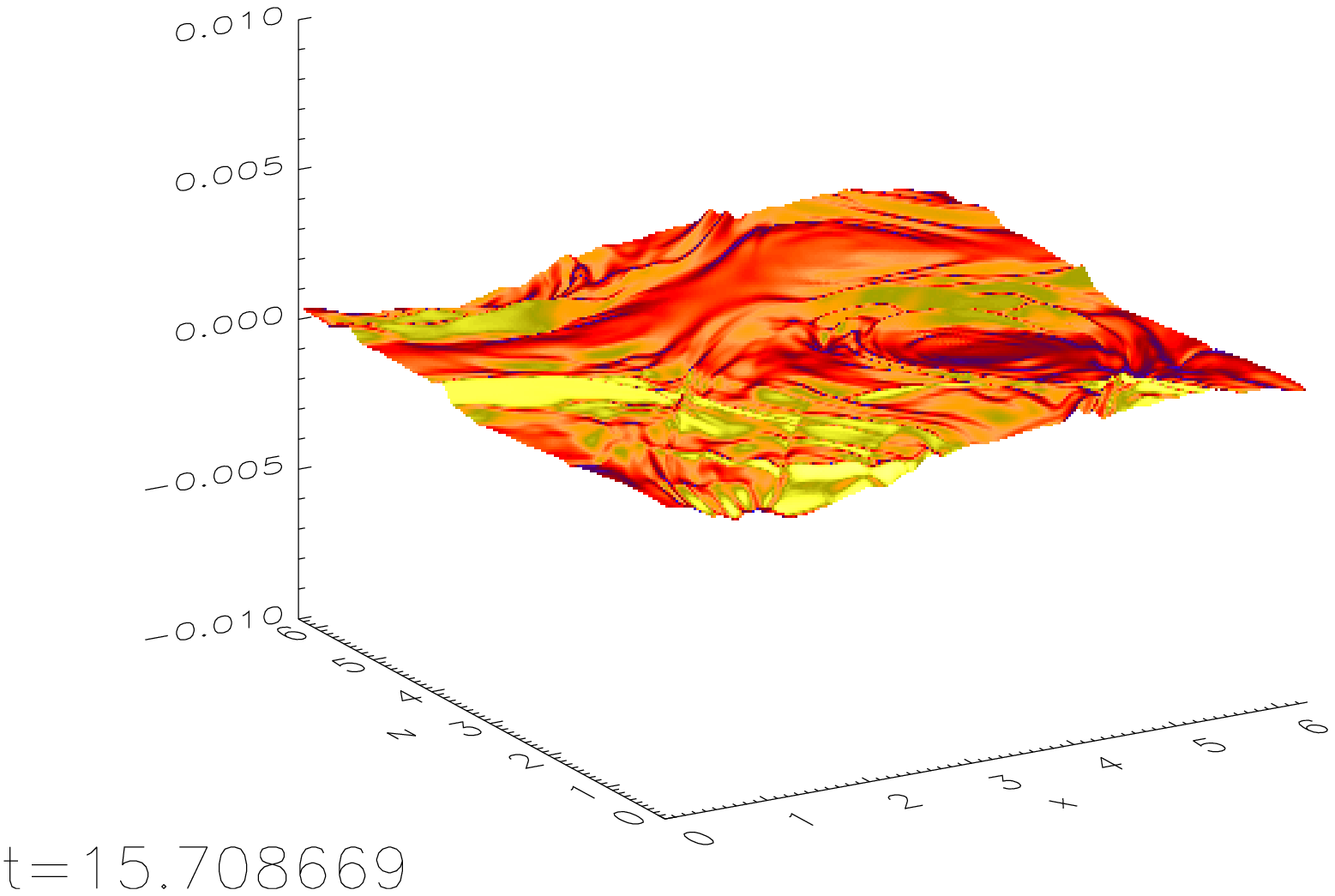}
\includegraphics[width=5.5cm]{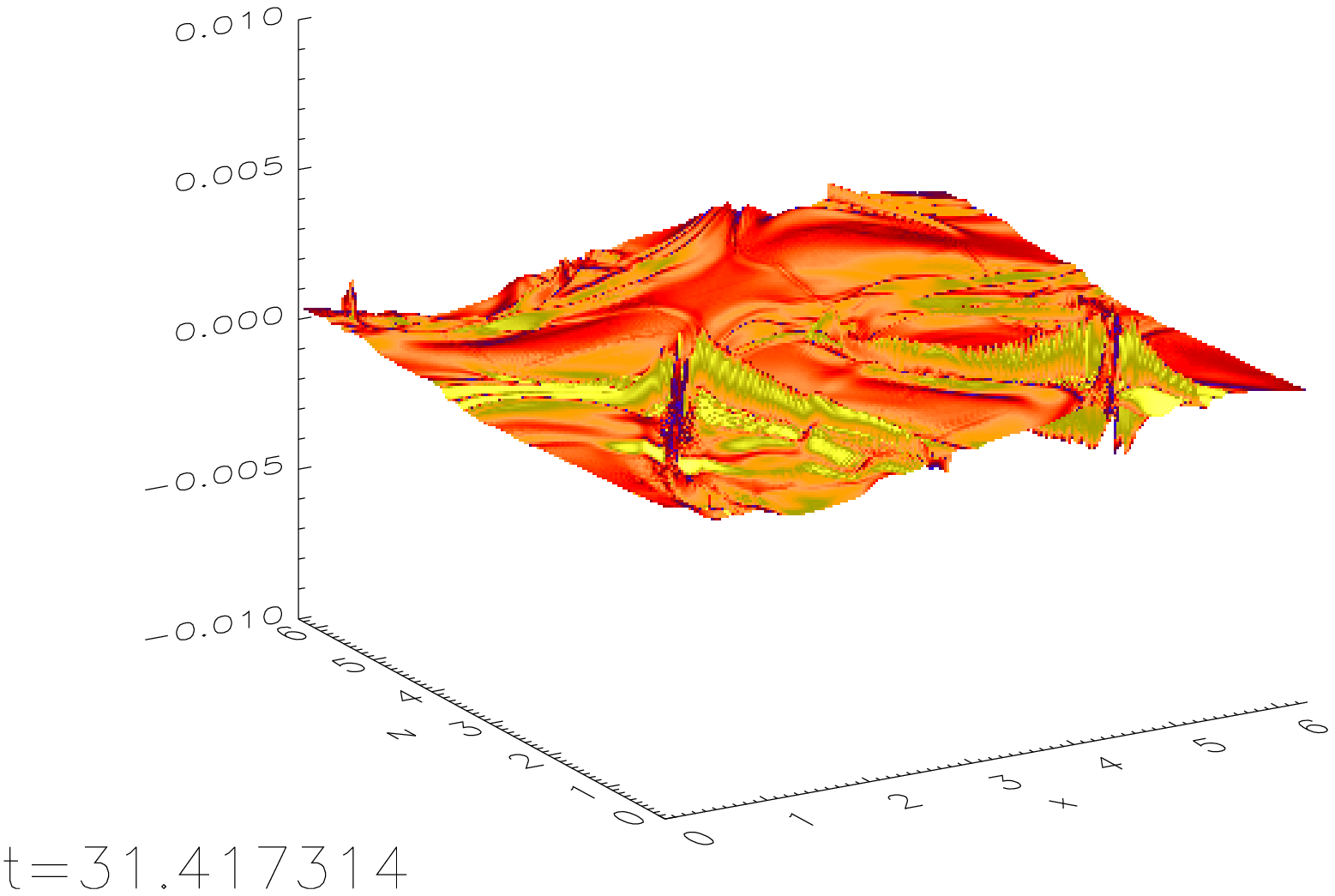}\\
\includegraphics[width=5.5cm]{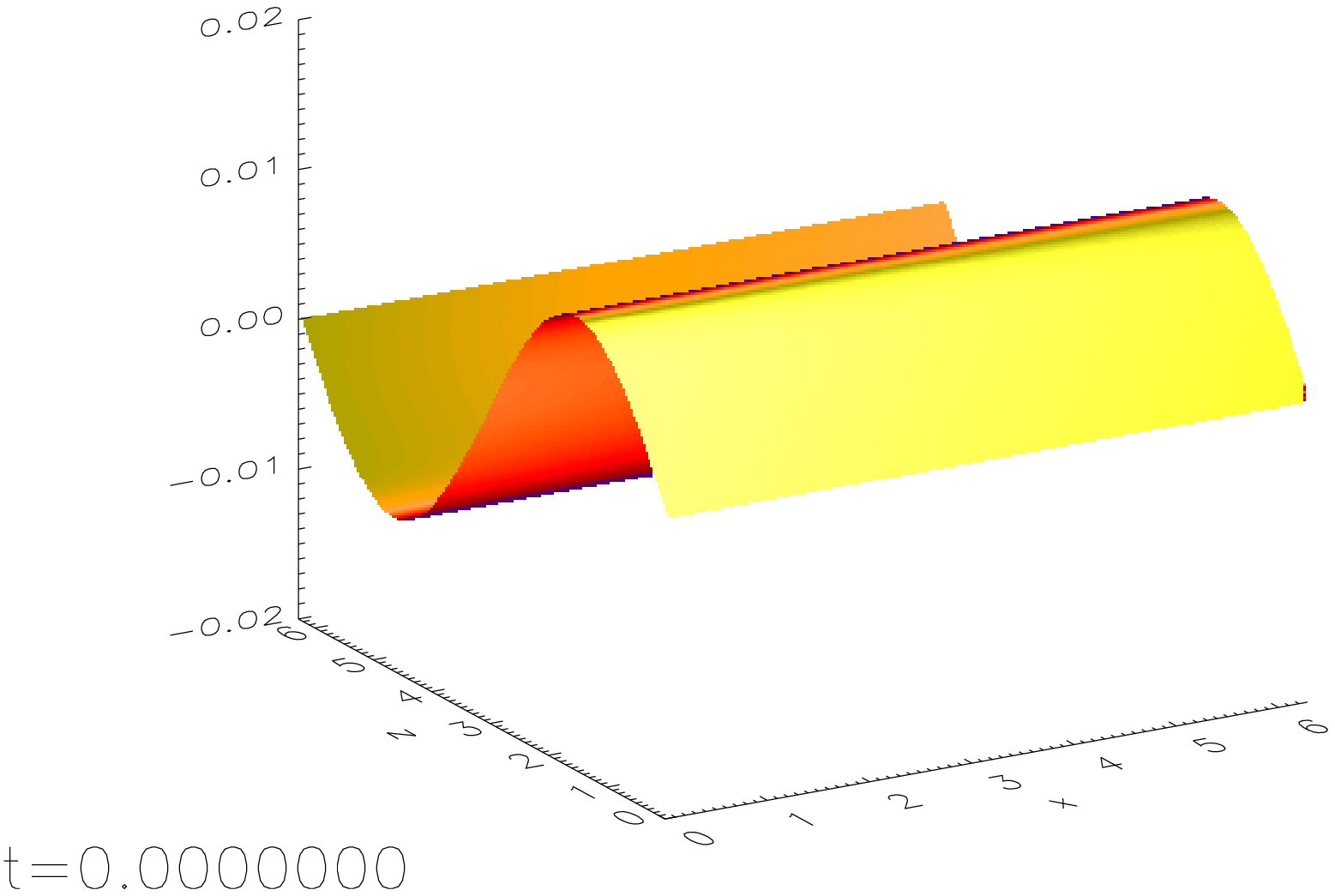}
\includegraphics[width=5.5cm]{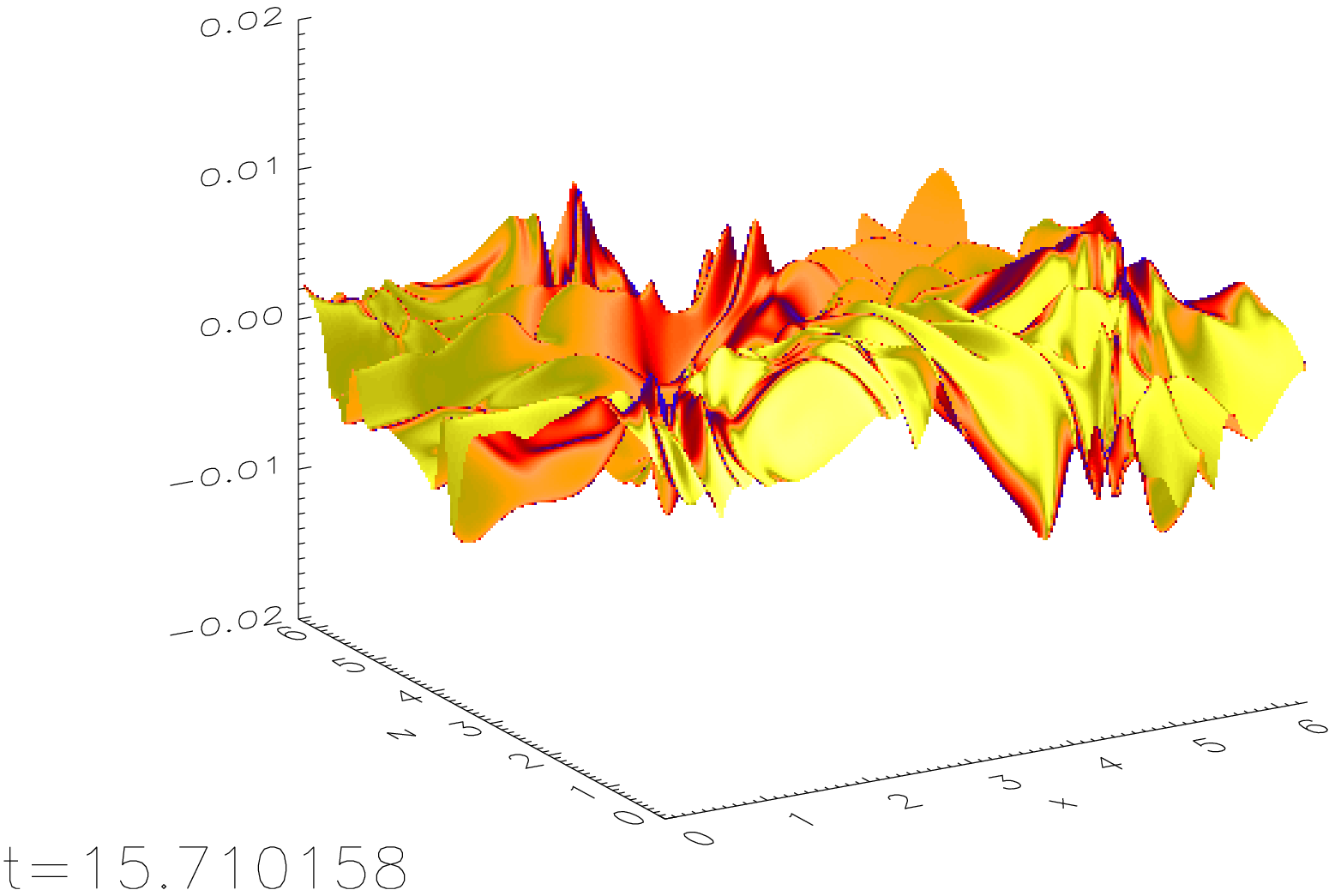}
\includegraphics[width=5.5cm]{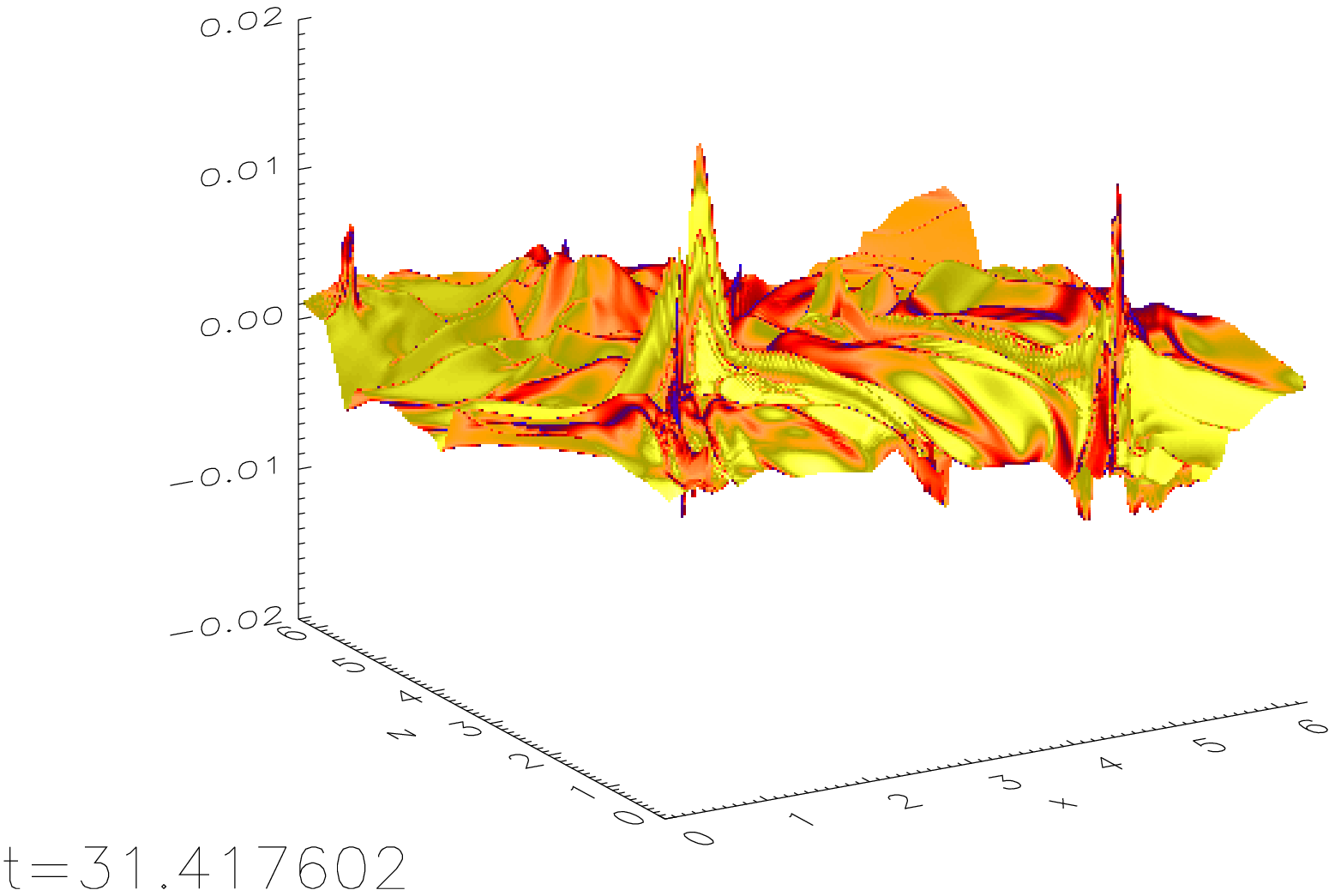}
\end{center}
\caption{Top panels are three time snapshots $t=0,t=t_{end}/2,t=t_{end}$ of 
$B_y(x,y=y_{max}/2,z)-B_{y0}(x,y=y_{max}/2,z,t)$ shaded surface plot,
i.e. difference between the magnetic field y-component in the case of ABC field with AW Gaussian pulse and with
resistivity $\hat \eta=5\times 10^{-4}$, i.e. numerical run $\mathrm{abc_p1}$, 
(denoted by $B_y(x,y=y_{max}/2,z)$ in the the animated version of this figure i.e. Movie 4 from Ref.\cite{mov}) 
and  magnetic field y-component in the case of ABC field without a AW pulse and with
the same resistivity $\hat \eta=5\times 10^{-4}$, i.e. numerical run $\mathrm{abc_e1}$, (denoted by $B_{y0}(x,y=y_{max}/2,z)$
in Movie 4).
Bottom panels are the same as top ones, but for the case of harmonic wave 
(the animated version of this figure's bottom row is shown in Movie 5 from Ref.\cite{mov}), 
therefore now corresponding numerical runs used in the subtraction are 
$\mathrm{abc_h1}$ and $\mathrm{abc_e1}$.}
\label{fig4}
\end{figure*}

\begin{figure*}[htbp] 
\begin{center}
\includegraphics[width=16.5cm]{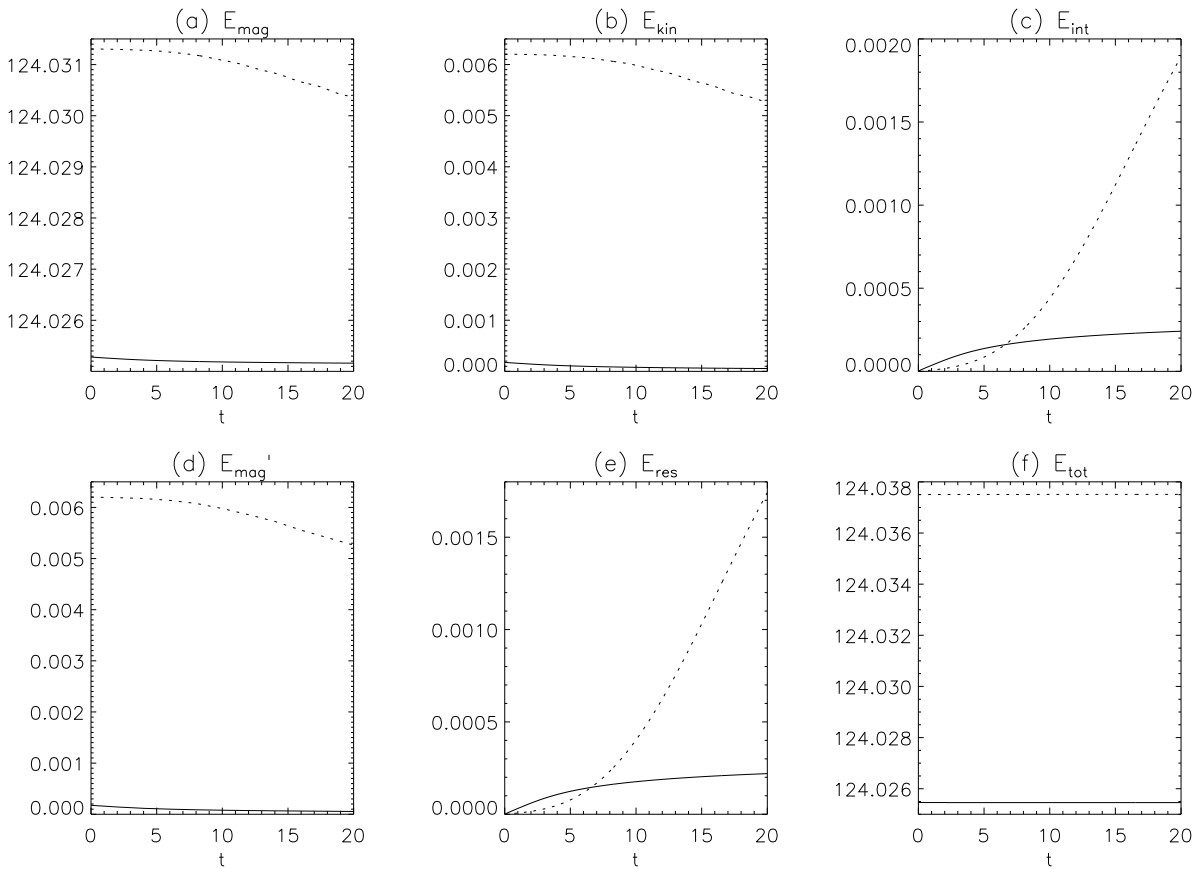}
\end{center}
\caption{Energies for the case of UBMF integrated over entire simulation box. 
In particular, the panels show: 
(a) full magnetic energy (background plus AW perturbation), $E_{mag}$, 
(b) full kinetic energy (background plus AW perturbation), $E_{kin}$, 
(c) internal energy, $E_{int}$, 
(d) magnetic perturbation energy, according to Eq.(5), $E_{mag}^{\prime}$,
(e) resistive energy (Ohmic heating), $E_{res}$,
(f) total energy,
$E_{tot}\equiv E_{mag}+E_{kin}+E_{int}$.
In all panels solid line is for run 
$\mathrm{con_p}$, while dotted line is for $\mathrm{con_h}$.}
\label{fig5}
\end{figure*}

\begin{figure*}[htbp] 
\begin{center}
\includegraphics[width=16.5cm]{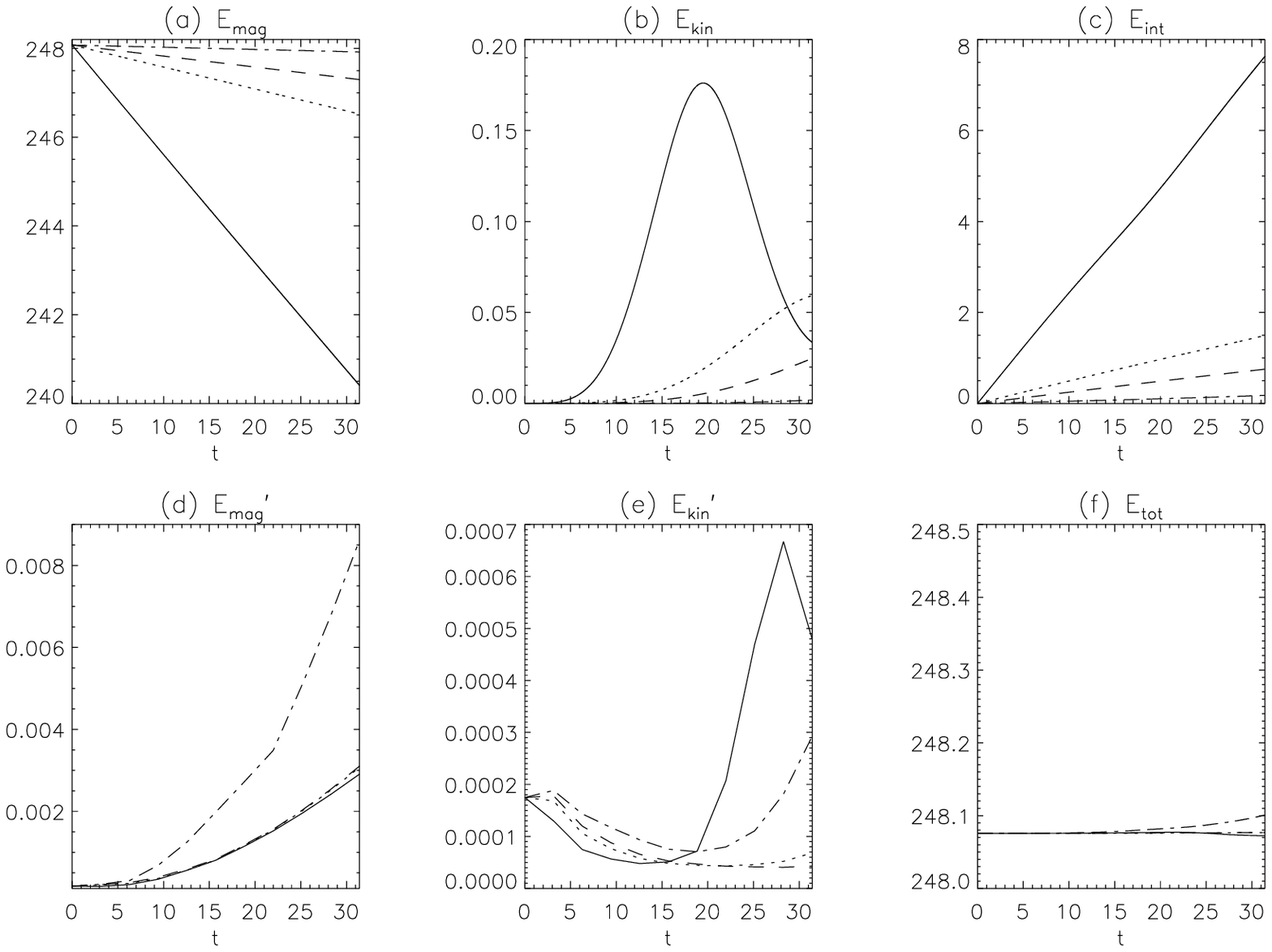}
\end{center}
\caption{Similar to Fig.~5 but here
 solid, dotted,  dashed, dash-dotted lines pertain to the runs
 $\mathrm{abc_p1}$, $\mathrm{abc_p2}$, $\mathrm{abc_p3}$ and $\mathrm{abc_p4}$ 
 respectively and panel (e) here is replaced by 
 kinetic perturbation energy,  $E_{kin}^{\prime}$, according to Eq.(6).}
\label{fig6}
\end{figure*}

\begin{figure*}[htbp] 
\begin{center}
\includegraphics[width=16.5cm]{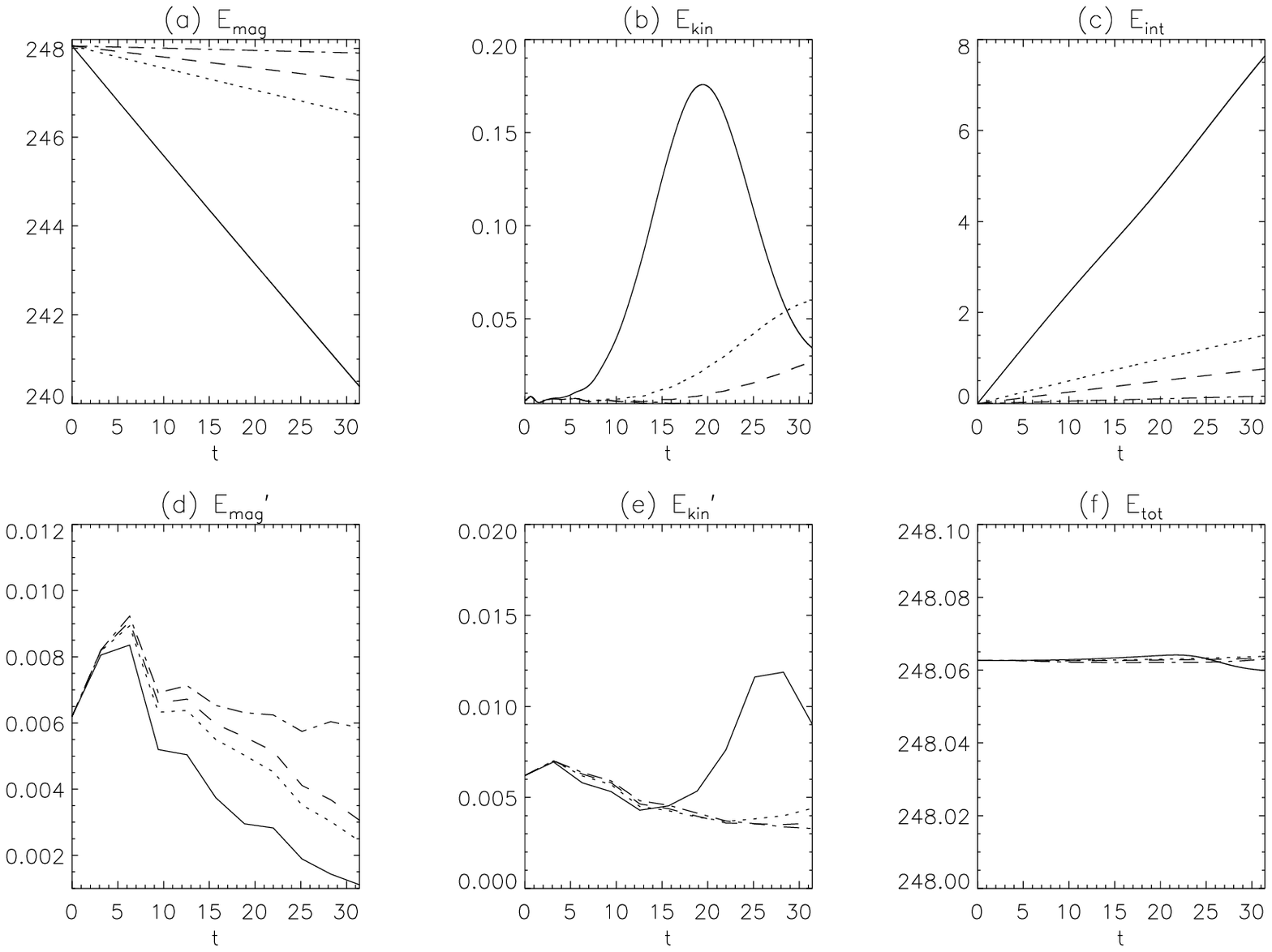}
\end{center}
\caption{Similar to Fig.~5 but here
 solid, dotted,  dashed, dash-dotted lines pertain to the runs
 $\mathrm{abc_h1}$, $\mathrm{abc_h2}$, $\mathrm{abc_h3}$ and $\mathrm{abc_h4}$ 
 respectively and panel (e) here is replaced by 
 kinetic perturbation energy,  $E_{kin}^{\prime}$, according to Eq.(6). }
\label{fig7}
\end{figure*}

Although Lare3d has been extensively tested before, there were several significant recent updates.
Thus, we start from presenting numerical code validation. This is done showing the results from 
two numerical runs of the code. In all our numerical simulations we use
 a 3D box with 512$^3$ uniform grids in $x$,$y$, and $z$
direction having length of $2\pi$ in each  spatial direction. 
Distance, magnetic field and density are normalised to their background values $L_0,B_0,\rho_0$.
Whereas, velocity and time to the respective Alfven values 
$C_A=B_0/\sqrt{\mu_0 \rho_0}$ and $\tau_A=L_0/C_A$.
Boundary conditions are periodic in all three spatial directions.
When using resistivity we have tested cases with zero and non-zero values
of the resistivity in the ghost cells around the physical simulation domain.
No noticeable difference was found by setting resistivity to zero in the ghost cells.
For the first two runs, normalised, uniform magnetic field, of strength unity, is in $z$-direction.
Plasma density has a profile in $x$-direction $\rho(x)=1+9\exp(-(x-\pi)^4 )$. 
For the runs with ABC magnetic field (see below) the density is set constant $\rho=1$.
Plasma beta and 
gravity are set to zero in all numerical runs.
We launch: 
(i) a Gaussian pulse which has two components, 
$B_y= 0.01 \exp(-(z-0.5)^2/(2 \times 0.05^2))$,
$V_y= -0.01 \exp(-(z-0.5)^2/(2 \times 0.05^2))/\sqrt{\rho(x)}$,
making it a linearly polarised AW packet, 
which has an amplitude of $0.01$ (except for numerical runs
$\mathrm{abc_a1}$ where amplitude is $0.05$ and 
$\mathrm{abc_a2}$ where amplitude is $0.1$), starts at $z=0.5$ and has a width of $0.05$. 
(ii) a harmonic wave which has two components,
$B_y= 0.01 \sin(z)$,
$V_y= -0.01 \sin(z)/\sqrt{\rho(x)}$,
making it also a linearly polarised AW packet, with an amplitude of $0.01$, and
spanning the full domain length in $z$
direction (contrary to pulse case that is rather spatially localised).
Plasma viscosity is set to zero, while first and second shock viscosity coefficients are 0.01 and 0.05
(see Ref.\cite{2001JCoPh.171..151A} for further details). 
Figs.~\ref{fig1} and \ref{fig2} show time evolution  of the AW damping for plasma resistivity of 
$\hat \eta=5\times 10^{-4}$
for the cases of Gaussian pulse and harmonic wave respectively 
(runs $\mathrm{con_p}$ and $\mathrm{con_e}$ from Table~\ref{runs}).
The resistivity is in units of $\mu_0 L_0 C_A$. Thus, $1/ {\hat \eta} =S$ is the Lundquist number.
In the both figures, crosses and open diamonds are numerical simulation results in the
strongest density gradient point $x=(155/512)\times (2\pi)=1.90214$ and 
away from the gradient $x=(1/512)\times (2\pi)=0.0122718$ (first grid cell in $x$-direction).
We essentially plot the simulation values by tracing crests of the  
numerical arrays $B_y(155,256,z)$ and $B_y(1,256,z)$ that
track damping of the AW.
In Fig.~\ref{fig1} the thin solid line corresponds to the asymptotic solution for large times
\begin{equation}
B_y=t^{-3/2}/(5 \sqrt{2\pi C^\prime_A(x)^2/3}),
\end{equation}
i.e. true for $t \gg \tau_A$, while a more general analytical form \cite{2002RSPSA.458.2307W,2003A&A...400.1051T}
\begin{equation}
B_y={\frac{\alpha_0}{ \sqrt{1+ 
\eta C^\prime_A(x)^2 t^3/ 3 \sigma^2}}} \\
\exp{\left[{-{\frac{(z-C_A(x) t)^2}{2 (\sigma^2+
\eta C^\prime_A(x)^2 t^3/ 3)}}}\right]}, 
\end{equation} 
is plotted with stars connected by thick line.
In Fig.~\ref{fig2} stars connected by thick line correspond to the analytical solution \cite{1983A&A...117..220H}
\begin{equation}
B_y=\exp(-\eta C^\prime_A(x)^2 t^3 k^2/6)\exp(-ik(z-C_A(x)t)).
\end{equation}
Note in Fig.~\ref{fig2} that away from the density gradient AW is 
not damped noticeably (open diamonds near top of the figure).
We see that AW damping closely follows the analytical theory expressions. The percentage errors at $t= 20$ 
Alfven times are 2.6\% for the 
case of Gaussian pulse and 1.1\% in the case of harmonic wave.
Naturally, the pulse is more spatially localised while the harmonic 
wave spans entire domain length in $z$ direction, thus when resolved by 512
grid points the larger error is for the case of the Gaussian pulse.
Movies 1 and 2 from Ref.\cite{mov} show time evolution of AW damping in the Gaussian pulse and harmonic
wave respectively. Note how AWs quickly damp (contours thin or fade away) in the density
inhomogeneity regions $x\approx1.5-2$ and $x\approx4.0-4.5$ where 
the wave fronts bend strongly starting at $t=0$ from
initially flat profiles. It is this derivative of the Alfven speed, $C^\prime_A(x)$,
in $x$-direction, which enters equations (1)--(3), that is responsible for the fast damping of the AW.
Away from the density gradient regions much slower
damping, $\exp(-\eta k^2 t)\propto \exp(-t)$, operates, 
which is barely noticeable on the time-scales concerned.

It should be noted we have checked how well the resistive equilibrium holds in the case of
UBMF. This was done in runs $\mathrm{con_e}$  from Table~\ref{runs}. We confirm that e.g.
the difference of magnetic field component $B_y(x,y=y_{max}/2,z,t=t_{end}=20)$ with its initial value
$B_y(x,y=y_{max}/2,z,0)$ with $\hat \eta=5\times 10^{-4}$ 
does not exceed $3\times10^{-14}$, i.e. resistive equilibrium holds
and subtracting the modification of $B_y(x,y=y_{max}/2,z,t)$ from the ideal 
(non-resistive) plasma case does not make any difference.  
However, the same does not hold for the ABC field and the resistive evolution of the
background magnetic field turns out to be significant.

\begin{figure}[htbp] 
\begin{center}
\includegraphics[width=8.5cm]{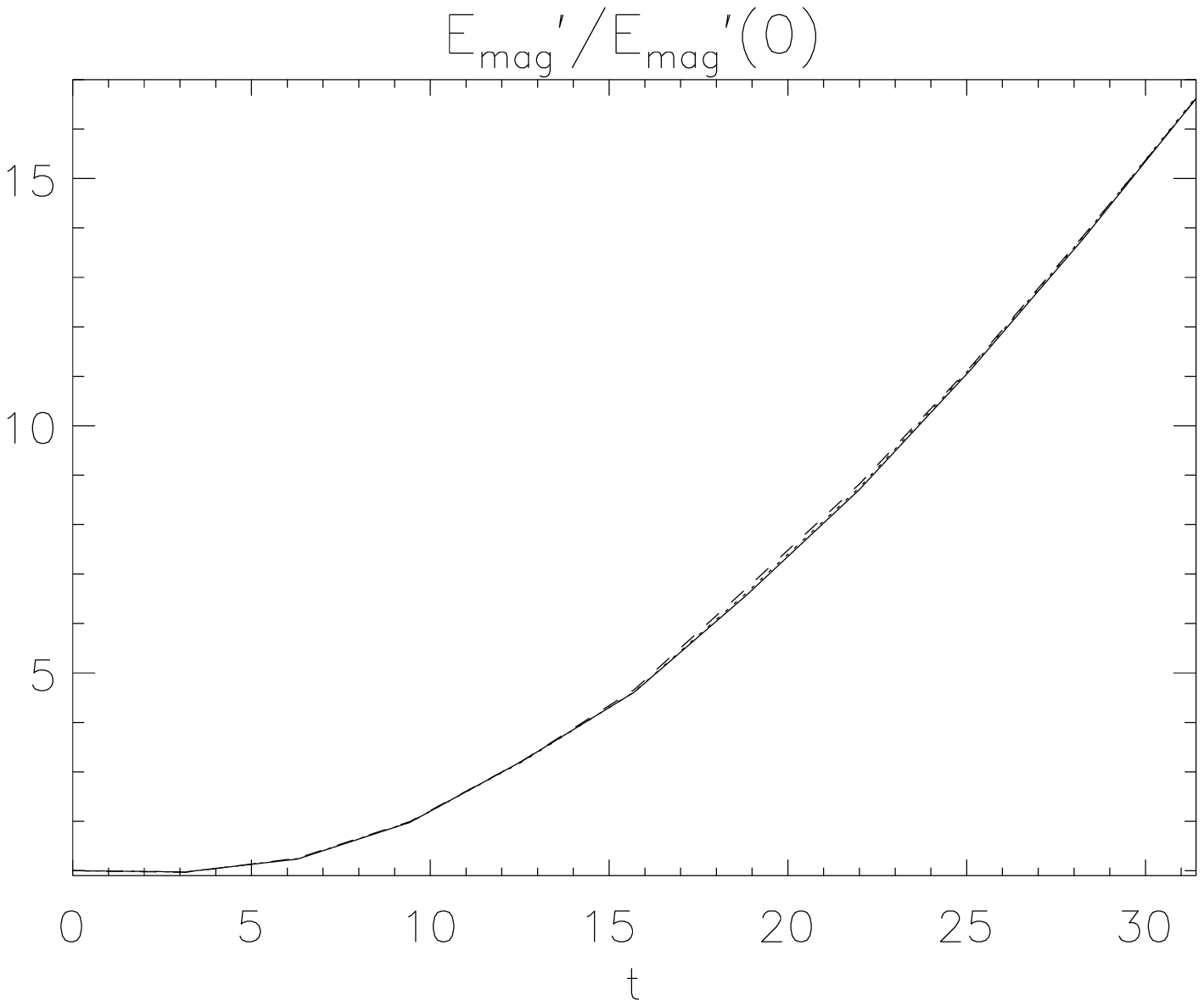}
\end{center}
\caption{Magnetic perturbation energy, normalised to its initial value, according to Eq.(5), 
$E_{mag}^{\prime}/E_{mag}^{\prime}(0)$,
for runs $\mathrm{abc_p1}$ (with the pulse amplitude $0.01$) solid line, 
$\mathrm{abc_a1}$ (with pulse amplitude $0.05$) dotted line and 
$\mathrm{abc_a2}$ (with pulse amplitude $0.1$)
dashed line.}
\label{fig8}
\end{figure}

\begin{figure}[htbp] 
\begin{center}
\includegraphics[width=8.5cm]{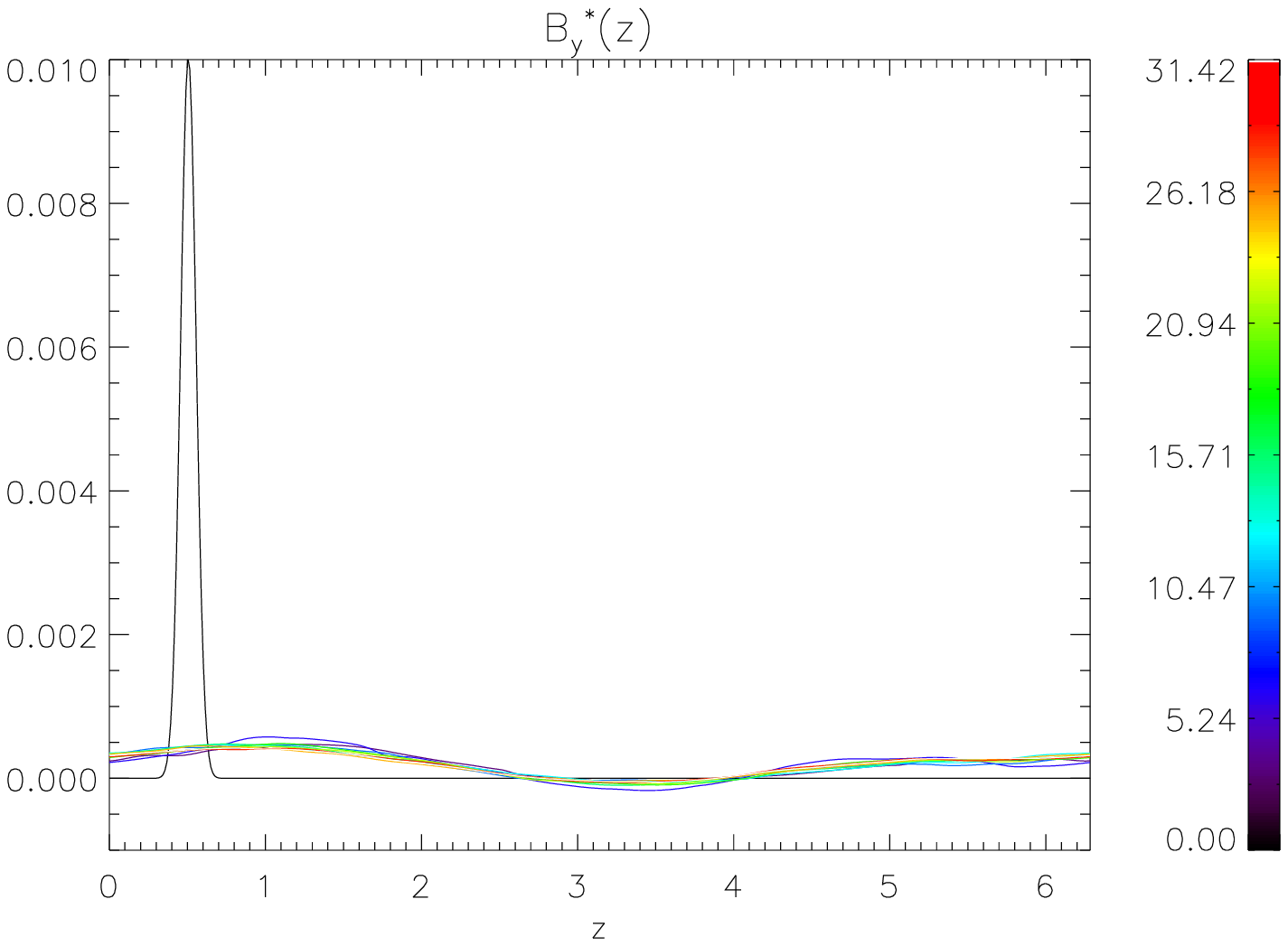} \\
\includegraphics[width=8.5cm]{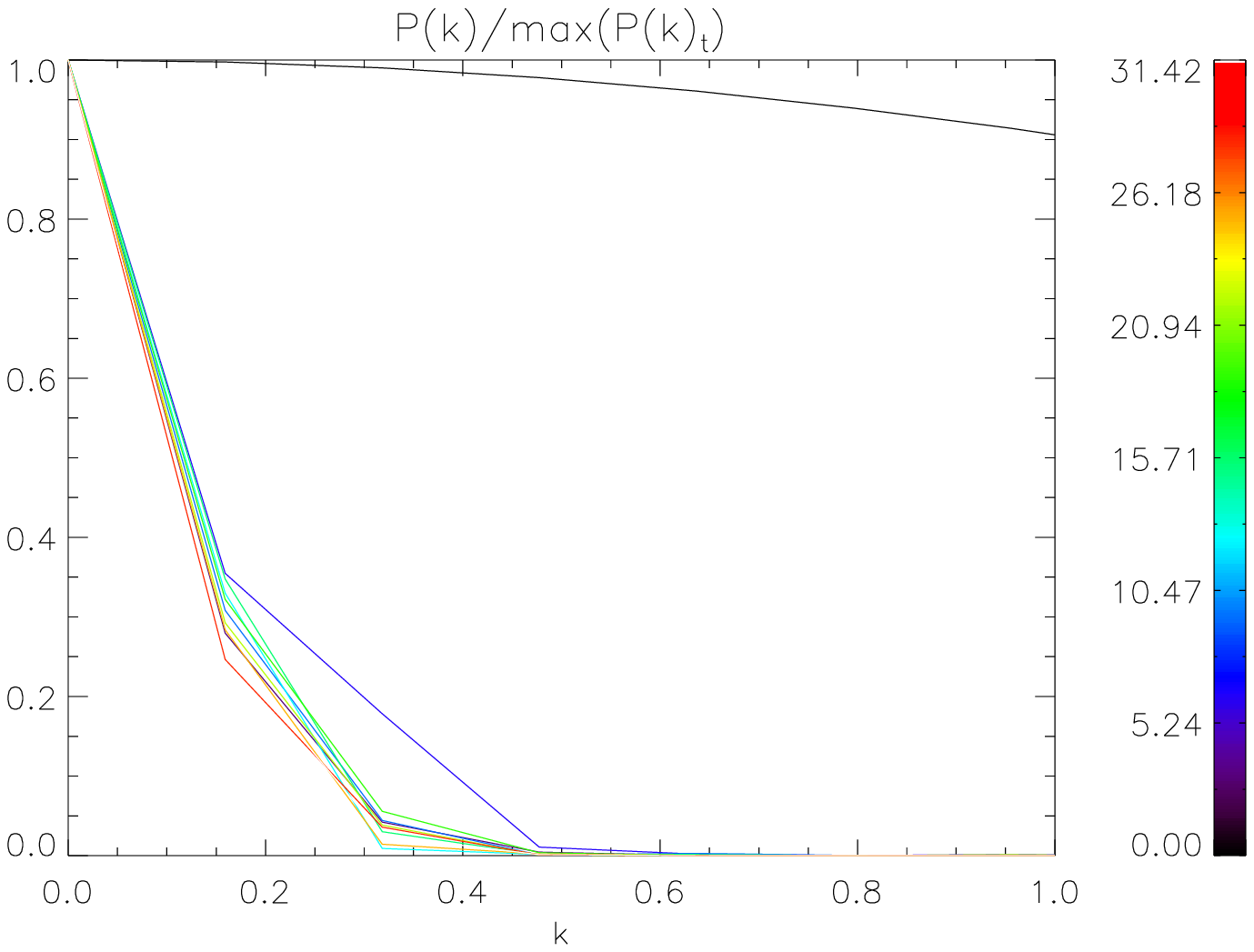}
\end{center}
\caption{Top panel: time evolution of
physical quantity $B_y^*(z)$ calculated by Eq.(7) for the case of Gaussian AW pulse.
The numerical runs used in the subtraction are 
$\mathrm{abc_p1}$ and $\mathrm{abc_e1}$.
Bottom panel: time evolution of the Fourier spectrum of $B_y^*(z)$ from the top panel.
See text for the normalisation of the Fourier spectrum used.
Colour bar and colour lines show the advance of simulation time 
from $t=0$ (black) to $10\pi$ (red). }
\label{fig9}
\end{figure}

\begin{figure}[htbp] 
\begin{center}
\includegraphics[width=8.5cm]{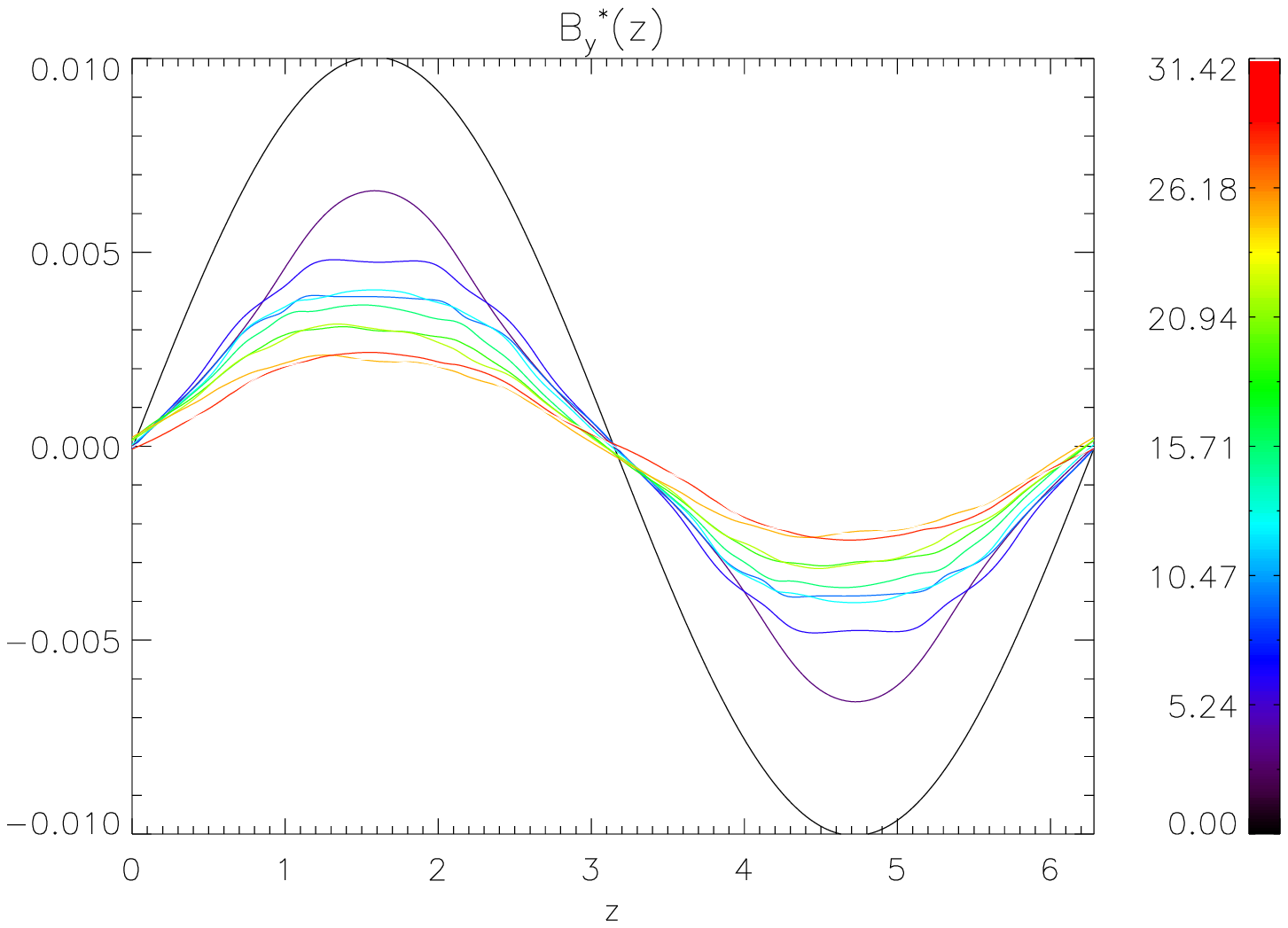} \\
\includegraphics[width=8.5cm]{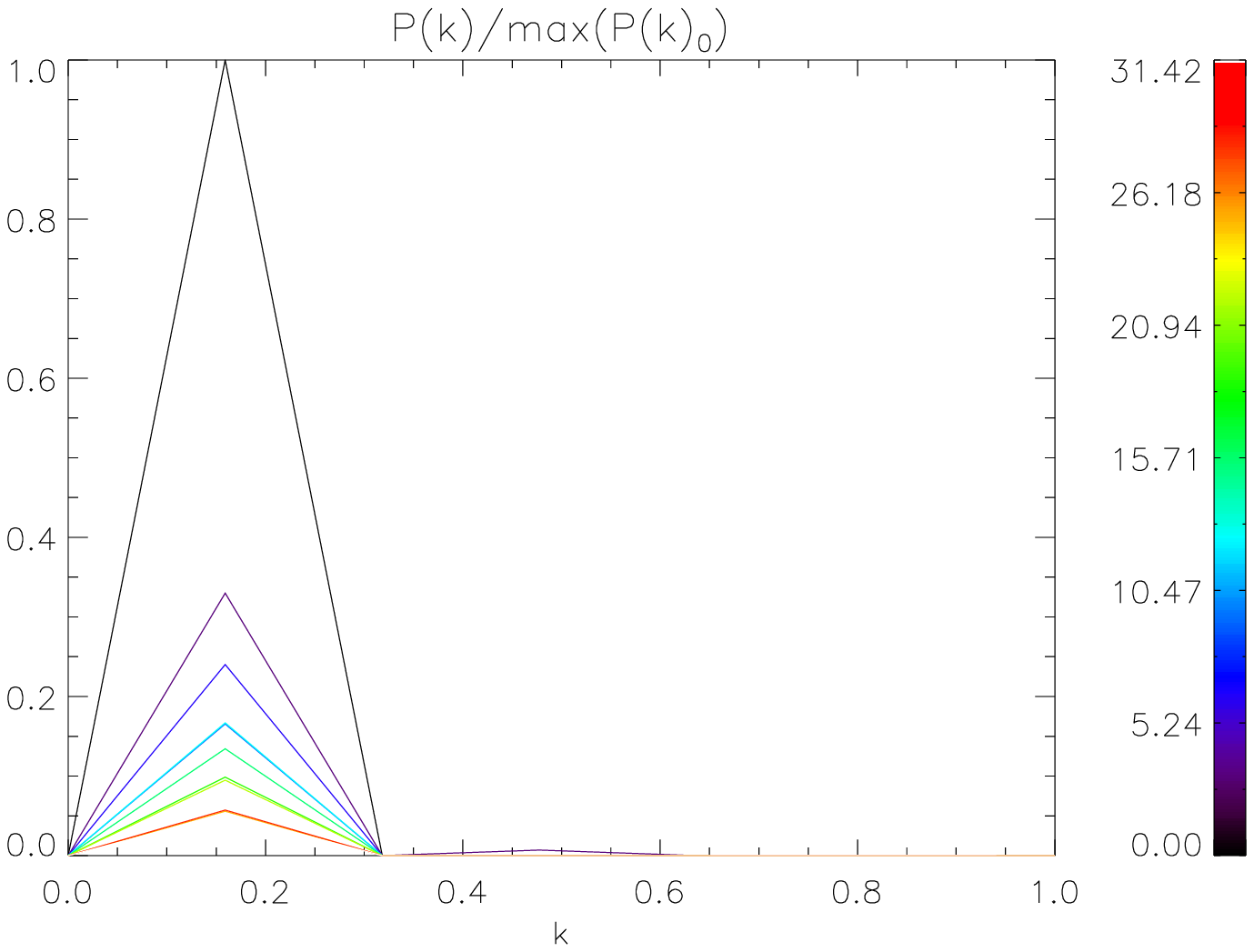}
\end{center}
\caption{The same as in Fig.~\ref{fig9} but for the case of the harmonic AW.
The numerical runs used in the subtraction here are 
$\mathrm{abc_h1}$ and $\mathrm{abc_e1}$.}
\label{fig10}
\end{figure}

The ABC magnetic field is given by the following expressions 
\cite{2000ApJ...533..523M}:
\begin{eqnarray}
B_x(x,y,z)=A\sin(z)+C\cos(y),&   \nonumber \\
B_y(x,y,z)=B\sin(x)+A\cos(z),&  \nonumber \\
B_z(x,y,z)=C\sin(y)+B\cos(x).& \nonumber \\
\end{eqnarray}
Following Ref.\cite{2000ApJ...533..523M} we fix the values of the coefficients as
$A=\sqrt{1/3}, \, B=1, \, C=\sqrt{2/3}$. This choice insures that ABC field
has essentially entangled magnetic flux tube-like structure along $z$-coordinate,
with regions of space that have regular (nearly uniform) magnetic flux-tubes
and also regions that have exponentially divergent magnetic fields (cf. Fig.~1
from Ref.\cite{2000ApJ...533..523M}). It is easy to show by an analytical calculation that
in the ideal ($\hat \eta=0$) case for the ABC field the usual plasma MHD equilibrium equation
$\nabla(p+ \vec B^2/(2 \mu_0))=(\vec B \cdot \nabla) \vec B)/\mu_0$
holds when pressure $p$ is constant (or zero).
However, the latter equation ignores the resistive effects and when included
these drive the magnetic field out of equilibrium.
The deviation from the initial equilibrium is studied in Fig.~\ref{fig3} and its animated
version Movie 3 from Ref.\cite{mov}, where we plot 
time dynamics of $B_y(x,y=y_{max}/2,z,t)-B_{y0}(x,y=y_{max}/2,z,t)$ shaded surface plot,
i.e. difference between the magnetic field y-component in the case of ABC field without  AW pulse but with
resistivity $\hat \eta=5\times 10^{-4}$, i.e. numerical $\mathrm{abc_e1}$, 
(denoted by $B_y(x,y=y_{max}/2,z)$ in Movie 3) 
and  magnetic field y-component in the case of ABC field without  AW pulse and without
resistivity $\hat \eta=0$, i.e. numerical $\mathrm{abc_e0}$, (denoted by $B_{y0}(x,y=y_{max}/2,z)$ in Movie 3).
It can be deduced that the difference attains a value of 0.029228 which is about three times the amplitude
of the AW (either Gaussian pulse or harmonic wave).
Other ($x$- and $z$-) magnetic field component differences are of the same order. 
It should be noted that the difference scales with $\hat \eta$, i.e.
in the run $\mathrm{abc_e3}$ where $\hat \eta=5\times 10^{-5}$ the difference is 0.0029228.

Therefore it is rather important {\it to take into account resistive evolution} of the
equilibrium of the magnetic field when studying propagation of AWs in ABC fields.
In practice this means that when launching AWs {\it we need to correctly single out
the magnetic perturbation from the background, i.e. ABC field without an AW pulse but with
the same resistivity}. This is achieved by calculation using Eq.(5), where
$\vec B$  (and its components $B_x$, $B_y$ and $B_z$) stands for full magnetic field (background plus AW) while $\vec B_0$
stands for just background i.e. ABC field without an AW pulse but with
the same resistivity.
\begin{eqnarray}
E_{mag}^{\prime}={\frac{1}{2}}\int_0^{2\pi}\int_0^{2\pi}\int_0^{2\pi}
[(B_x^2+B_y^2+B_z^2)-&   \nonumber \\
(B_{x0}^2+B_{y0}^2+B_{z0}^2)-2 \vec B_0 \cdot (\vec B-\vec B_0)
]dxdydz.
\end{eqnarray}
A similar approach is adopted for the velocity perturbations:
\begin{eqnarray}
E_{kin}^{\prime}={\frac{1}{2}}\int_0^{2\pi}\int_0^{2\pi}\int_0^{2\pi}
[(V_x^2+V_y^2+V_z^2)-&   \nonumber \\
(V_{x0}^2+V_{y0}^2+V_{z0}^2)-2 \vec V_0 \cdot (\vec V-\vec V_0)
]dxdydz.
\end{eqnarray}
As with magnetic field in Eq.(6), 
$\vec V$  (and its components $V_x$, $V_y$ and $V_z$) 
stands for full velocity field (background plus AW) while $\vec V_0$
stands for just background i.e. ABC field without an AW pulse but with
the same resistivity.

The dynamics of AW Gaussian pulse and harmonic wave is shown in Fig.~\ref{fig4}
with corresponding animated versions presented in Movie 4 and Movie 5 from Ref.\cite{mov}.
It is clear that both AWs damp rather quickly. Note that both in Fig.~\ref{fig1}, Fig.~\ref{fig2},
and Fig.~\ref{fig4} resistivity is the same ($\hat \eta=5 \times 10^{-4}$), and the end simulation times
are $t_{end}=20$ and $10\pi$, respectively. This shows that in the case of ABC 
background field the AW damping is faster. Note from the movies that, because of existence of 
exponentially divergent field line regions, initially flat AW fronts became quickly corrugated
via wave refraction, 
because of rather complicated form of the local Alfven speed $C_A(x,y,z)$ prescribed by Eq.~(4).
As in the case of UBMF, in ABC fields the Gaussian pulse damping is also 
faster because its strong localisation along
$z$-coordinate.

The most interesting findings of this study come to light when investigating the
energetics of the AW dynamics/damping.
Fig.~\ref{fig5} shows energies for the UBMF.
The energies are calculated over entire simulation domain using
Lare3d code's built-in function called $\mathrm{getenergy()}$.
The latter takes the averages of $B_x^2/2$, $B_y^2/2$ and $B_z^2/2$
to cell centres and then sums over all simulation cells.
The latter function produces magnetic, kinetic, internal and resistive (Ohmic) heating energies.
The only exception is panel (d) where magnetic perturbation energy is plotted, 
according to Eq.(5).
Note that in panels (d), in all figures \ref{fig5}, \ref{fig6}, \ref{fig7} and in Fig.~\ref{fig8},
there are only 10 equally spaced data points, while in all other panels there thousands of data points.
This is because Lare3d's $\mathrm{getenergy()}$ outputs energy at every time step, whereas
when we use Eqs.(5) and (6), we use IDL's built in function $\mathrm{int\_tabulated}$,
which employs five-point Newton-Cotes integration formula,
to do the manual integrations. We have tested energy calculation
using $\mathrm{int\_tabulated}$ and $\mathrm{getenergy()}$ and concluded while the
both yield similarly close results, $\mathrm{int\_tabulated}$ has a superior accuracy.

One can  gather from Fig.~\ref{fig5}(a,b,d) that the magnetic, kinetic 
and AW perturbation magnetic energies 
start from their respective values and then decrease in time. 
Whereas internal  and resistive heating energies start from zero and increase in time -- 
see Fig.~\ref{fig5}(c,e).
Note that all of the above energies include contributions from both
inhomogeneous density parts where AW damping is rather vigorous and homogeneous density
parts where damping is weak. Because density gradient regions are not large, overall AW 
damping is not strong. Overall, UBMF cases produce the expected result that AW perturbation energy is damped
and converted in plasma resistive heating. 
We see in Fig.~\ref{fig5}(f) that the total energy
is conserved, indicating that numerical errors (numerical dissipation) are tolerably small.

The most surprising result is obtained in the study of ABC-field energetics, shown
in Fig.~\ref{fig6} and Fig.~\ref{fig7}.
The latter two plots are similar to Fig.~\ref{fig5}, but now show energies for the ABC background
magnetic field for the Gaussian pulse and harmonic wave cases, respectively.
The four curves  in Fig.~\ref{fig6} and Fig.~\ref{fig7} correspond to the 
four different resistivities, as detailed in Table~\ref{runs}.

One can  gather from Fig.~\ref{fig6}(a) and Fig.~\ref{fig7}(a) that the total
magnetic energies  start from their respective initial values and then decrease in time.
Note that, naturally, larger resistivity results in larger dissipation of magnetic energy.
The total kinetic energy seems to transiently increase, see solid curves in 
Fig.~\ref{fig6}(b) and Fig.~\ref{fig7}(b),
and this can be attributed to the resistive evolution of the background.
The transient increase in the total kinetic energy can be only seen for 
$\hat \eta=5\times 10^{-4}$. It is not certain, but is a likely 
possibly that for smaller resistivity (dotted, dashed and dash-dotted curves in 
Fig.~\ref{fig6}(b) and Fig.~\ref{fig7}(b)) will behave in a similar time-transient way.

Note that Figs.~\ref{fig6}(a), (b) and (c) are nearly identical to
Figs.~\ref{fig7}(a), (b) and (c), with the exception of Fig.~\ref{fig7}(b),
where a small "wiggle" in the lower left corner is noticeable.
This similarity can be attributed to the fact that the behavior of
the energies  is prescribed by the resistive evolution of the ABC
background magnetic field rather than AW perturbation damping.
The proof of this can be found by looking at 
Figs.~\ref{fig5}(a), (b) and (c) where cases with UBMF are shown.
Here the corresponding energies behave distinctly differently
in the case of different types of AW perturbations, because
now their time evolution is prescribed by AW damping and
there is {\it no} resistive evolution of the {\it uniform}
background magnetic field. 

Because in panels Fig.~\ref{fig6}(d) and Fig.~\ref{fig7}(d) behaviour
of $E_{mag}^{\prime}$ becomes different (at least on the timescales considered),
let us estimate the resistive times for the Gaussian and harmonic AWs.
Based on the one dimensional 
diffusion equation, $\partial B/\partial t=
{\hat \eta}\partial^2 B /\partial^2_{xx}$, which governs
phase-mixed AW damping, we define the 
resistive time as $\tau_r=L^2/{\hat \eta}$, where reciprocal 
of $1/L \approx \partial / \partial x$ is the length-scale of variation
of magnetic field in AW. Therefore, for the largest value of resistivity 
considered, $\hat \eta=5\times 10^{-4}$, we have 
$\tau_r=(2\pi)^2/5\times 10^{-4}=7.9 \times 10^4 \tau_A$ for the harmonic AW.
For Gaussian pulse case for the same resistivity 
$\tau_r=(0.05)^2/5\times 10^{-4}=5 \tau_A$. 
Here, Gaussian pulse width is taken as 0.05 which can be also inferred from the 
black solid curve in the top panel of Fig.9.
We gather from Fig.~\ref{fig6}(d) and Fig.~\ref{fig6}(e) that 
in the case of the Gaussian AW pulse the velocity perturbation energy 
growth is transient in time for $\hat \eta=5\times 10^{-4}$, 
attaining a peak within few resistive times $\tau_r=5$, while magnetic perturbation energy
continues to grow. The numerical values that can be also read from Fig.~\ref{fig6}(d) and Fig.~\ref{fig8}
are such that $E_{mag}^{\prime}(t=t_{end})/E_{mag}^{\prime}(0)=49.25$ for
$\hat \eta=10^{-5}$ (dash-dotted curve) and $E_{mag}^{\prime}(t=t_{end})/E_{mag}^{\prime}(0)=16.62$ for
$\hat \eta=5\times 10^{-4}$ (solid curve).
We also gather from Fig.~\ref{fig7}(d) and Fig.~\ref{fig7}(e) that for $\hat \eta=5\times 10^{-4}$,
in the case of the harmonic AW, the perturbation energy growth is transient in time, 
attaining peaks in both velocity ($t \approx 25$) and magnetic ($t \approx 5$)
perturbation energies within timescales much smaller than the resistive time 
$\tau_r=7.9 \times 10^4$.
The numerical values that can be also read from Fig.~\ref{fig7}(d)
are such that $max(E_{mag}^{\prime}(t)/E_{mag}^{\prime}(0))=1.49$ for
$\hat \eta=10^{-5}$  (dash-dotted curve) and $max(E_{mag}^{\prime}(t)/E_{mag}^{\prime}(0))=1.35$ for
$\hat \eta=5\times 10^{-4}$ (solid curve).
Figures with the  internal (Fig.~\ref{fig6}(c) and \ref{fig7}(c))
and resistive heating energies (not shown here) start from zero and increase in time.
Again larger resistivity results in the larger growth.
The unexpected result is that 
the magnetic perturbation energy, $E_{mag}^{\prime}$, calculated using Eq.(5),
increases in time, in the case of Gaussian pulse and time-transiently in the case of
harmonic AW, despite that (i) AW damps (cf. Fig.~\ref{fig4}) and 
(ii) total magnetic energy decreases in time (cf. Fig.~\ref{fig6}(a) and \ref{fig7}(a)).

The initial AW perturbation has $V_y$ and $B_y$ components and whilst the amplitudes are small,
0.01, one could conjecture that 
even a small non-linearity can produce a flow by a peristaltic mechanism 
\cite{1998JEnMa..34..435A}. As the AWs travel along the field lines
(due to the plasma frozen-in condition that is 
somewhat offset by a finite resistivity) the flow that derives 
from the AW perturbation might generate
the magnetic field by the dynamo action.
It is a well-known fact that flows that have similar mathematical
structure to Eq.(4) result in a magnetic dynamo action.
We explore this in Fig.~\ref{fig8} where essentially we repeat numerical run
$\mathrm{abc_p1}$, which has pulse amplitude of $0.01$, for two additional 
pulse amplitudes $0.05$ and $0.1$. Because the strength of the
peristaltic flow is an effect that is proportional to the amplitude
squared (i.e. quadratical non-linearity effect), one would expect a stronger growth
of AW magnetic perturbation energy with an increase of amplitude.
We gather from Fig.~\ref{fig8} that the increase of amplitude does not alter
AW magnetic perturbation energy growth. Thus both peristalsis and magnetic dynamo
action can be excluded as a cause of AW magnetic perturbation energy growth.

Next, we conjecture that magnetic perturbation energy growth can be attributed to the 
inverse cascade of
magnetic energy \cite{1967PhFl...10.1417K}. The conjecture of the inverse
cascade has strong support from both computer simulations
\cite{1986JFM...170..139S} and laboratory experiments \cite{2012PhRvE..85e6315B}.
We explore this idea in Fig.~\ref{fig9} for the case of the Gaussian AW pulse and 
in Fig.~\ref{fig10} for the case of the harmonic AW.
Colour bar and colour lines in both figures show advance of simulation time 
from $t=0$ (black) to $10\pi$ (red). Top panels in both figures show time evolution of
physical quantity $B_y^*(z)$ calculated by
\begin{eqnarray}
B_y^*(z)={\frac{1}{4\pi^2}}\int_0^{2\pi}\int_0^{2\pi}
(B_y-B_{y0})
dxdy,
\end{eqnarray}
where
 $B_y$ stands for full magnetic field (background plus AW) while $ B_{y0}$
stands for just background i.e. ABC field without an AW pulse but with
the same resistivity.
We clearly see at $t=0$ a Gaussian pulse with width 0.05 (Fig.~\ref{fig9}) and harmonic wave with
wavelength of $2\pi$ (Fig.~\ref{fig10}). Time evolution can be tracked by looking at different colour lines
which represent time interval of $\pi$ (ten of such intervals altogether).
We see in Fig.~\ref{fig9} that the Gaussian pulse quickly diffuses away by increasing its width.
In Fig.~\ref{fig10} we see that despite such complicated behaviour as seen in Fig.~\ref{fig4}, after all the
wave refraction, due to coordinate dependent Alfven speed is integrated out, the sinusoidal shape
is still retained and we only see the AW amplitude fading away. Note that the wave is not standing but 
moving many times in the periodic box -- it is the choice of snapshot times create this 
stroboscopic effect.
Bottom panel of Fig.~\ref{fig9} shows time evolution of the Fourier spectrum.
Each different colour line is normalised to a maximum value at different times (hence subscript in 
$max(P(k)_t)$) therefore all curves start from unity. Black curve that can be seen in the upper part of the
plot is actually a Gaussian, because Fourier transform of a Gaussian is a Gaussian. It appears as a very flat
Gaussian because we wanted to show clearly later time evolution, thus we had to restrict the range of wavenumbers
$k$ to unity. We see no evidence for the inverse cascade because no more wave power is seen at smaller $k$
for large times. We see a simple diffusion process of AW, when initially narrow Gaussian pulse widens by
diffusion (via resistivity). Bottom panel of Fig.~\ref{fig10} also 
shows time evolution of the Fourier spectrum but for harmonic AW. Now each different colour line is 
normalised to a maximum value at $t=0$ (hence subscript in 
$max(P(k)_0)$) therefore black curve peaks with the value of unity. At all later times the peak amplitudes
decrease because wave damps. The prominent peak is at $1/(2\pi)=0.159$, as expected of Fourier
transform of $\sin(z)$. More importantly, as the time progresses the peak does not shift in $k$, therefore again
we see no evidence for the inverse cascade.

\section{conclusions}
Motivated by Ref.\cite{2000ApJ...533..523M}, who studied small-amplitude 
AW packets in WKB approximation in ABC magnetic fields, we relax the approximation
and solve fully 3D MHD problem.
Ref.\cite{2000ApJ...533..523M} drew a distinction between 
2D AW dissipation via phase mixing, with AW dissipation time scaling of $S^{1/3}$, 
and 3D AW dissipation via exponentially divergent 
magnetic field lines, with dissipation time scaling of $\log(S)$.
They also suggested that for $S \leq 4 \times 10^6$ no clear distinction could be
drawn between the two regimes, as the large resistivity $\hat \eta =1/S=2.5 \times 10^{-7}$
made damping too strong. In the current study because we used full 3D MHD simulations 
(as opposed to WKB approximation used by Ref.\cite{2000ApJ...533..523M}) we could not
access large enough end simulation times for the damping to be noticeable in the small 
resistivity regime. Thus testing of the above AW damping scaling laws within full MHD
is still not achieved. However, we found other interesting effects:
We studied two types of AW perturbations: (i) a Gaussian pulse with length-scale much 
shorter than ABC domain length and (ii) a harmonic
AW with wavelength equal to ABC domain length.
We have shown that AWs dissipate quickly in the ABC field. 
Our results are surprising in that AWs magnetic perturbation energies increase in time, monotonously or
in time-transient manner, depending on the spatial scale of the AW disturbance, within
the considered end simulation time.
In the case of the harmonic AW the perturbation energy growth is transient in time, 
attaining peaks in both velocity and magnetic
perturbation energies within timescales much smaller than the resistive time.
In the case of the Gaussian AW pulse the velocity perturbation energy 
growth is also transient in time, 
attaining a peak within few resistive times, while magnetic perturbation energy
continues to grow.
We find that the total magnetic energy decreases 
in time and this is prescribed by the resistive evolution 
of the background ABC magnetic field rather than AW damping. 
Moreover, in the case of uniform background magnetic field, 
the total magnetic energy decrease in time is prescribed by AW damping, 
because of the absence of resistive evolution of the background.
We then considered runs with different amplitudes and performed 
analysis of the perturbation spectra. 
We excluded both (i) a possible dynamo action by AW perturbation-induced peristaltic flow and (ii) an 
inverse cascade of
magnetic energy. The only remaining reasonable explanation to the perturbation energy growth 
is a new instability.
The growth rate seems to be dependent of the value of the resistivity 
and also on the spatial scale of the AW disturbance. Further analysis is needed
in order to determine the exact mathematical nature of the growth rate dependence on these parameters.

The main conclusion is that in the complex, exponentially diverging  magnetic fields that can
occur e.g. in the lower solar corona, in cusps of Earth magnetosphere and/or Tokamak/Stellarator,
ABC-like background magnetic field, with periodic boundary conditions, evolve significantly in
time caused by the slow diffusion. The deviations from the initial state can be as large as 0.03
(with initial background magnetic fields being of the order of unity) within $10\pi\approx 30$ Alfven times.
Thus, the fast damping in these entangled magnetic fields, as predicted by the 
WKB approximation, seems not to be guaranteed.   

\begin{acknowledgments}
Author would like to thank (i) an anonymous referee, (ii) Prof. T.D. Arber (University of Warwick)
and (iii) Prof. S.M. Tobias (University of Leeds) for useful comments.
Computational facilities used are that of Astronomy Unit, 
Queen Mary University of London and STFC-funded UKMHD consortium at Warwick University. 
Author is financially supported by
STFC consolidated Grant ST/J001546/1,  Leverhulme
Trust Research Project Grant RPG-311 and
HEFCE-funded South East Physics Network (SEPNET).
\end{acknowledgments}


\begin{thebibliography}{20}
\expandafter\ifx\csname natexlab\endcsname\relax\def\natexlab#1{#1}\fi
\expandafter\ifx\csname bibnamefont\endcsname\relax
  \def\bibnamefont#1{#1}\fi
\expandafter\ifx\csname bibfnamefont\endcsname\relax
  \def\bibfnamefont#1{#1}\fi
\expandafter\ifx\csname citenamefont\endcsname\relax
  \def\citenamefont#1{#1}\fi
\expandafter\ifx\csname url\endcsname\relax
  \def\url#1{\texttt{#1}}\fi
\expandafter\ifx\csname urlprefix\endcsname\relax\def\urlprefix{URL }\fi
\providecommand{\bibinfo}[2]{#2}
\providecommand{\eprint}[2][]{\url{#2}}

\bibitem[{\citenamefont{{Aschwanden}}(2005)}]{2005psci.book.....A}
\bibinfo{author}{\bibfnamefont{M.~J.} \bibnamefont{{Aschwanden}}},
  \emph{\bibinfo{title}{{Physics of the Solar Corona. An Introduction with
  Problems and Solutions (2nd edition)}}} (\bibinfo{publisher}{Spinger-Praxis},
  \bibinfo{year}{2005}).

\bibitem[{\citenamefont{{Hung} and {Hassam}}(2013)}]{2013PhPl...20i2107H}
\bibinfo{author}{\bibfnamefont{C.~P.} \bibnamefont{{Hung}}} \bibnamefont{and}
  \bibinfo{author}{\bibfnamefont{A.~B.} \bibnamefont{{Hassam}}},
  \bibinfo{journal}{Physics of Plasmas} \textbf{\bibinfo{volume}{20}},
  \bibinfo{pages}{092107} (\bibinfo{year}{2013}).

\bibitem[{\citenamefont{{Podest{\`a}} et~al.}(2013)\citenamefont{{Podest{\`a}},
  {Gorelenkov}, {White}, {Fredrickson}, {Gerhardt}, and
  {Kramer}}}]{2013PhPl...20h2502P}
\bibinfo{author}{\bibfnamefont{M.}~\bibnamefont{{Podest{\`a}}}},
  \bibinfo{author}{\bibfnamefont{N.~N.} \bibnamefont{{Gorelenkov}}},
  \bibinfo{author}{\bibfnamefont{R.~B.} \bibnamefont{{White}}},
  \bibinfo{author}{\bibfnamefont{E.~D.} \bibnamefont{{Fredrickson}}},
  \bibinfo{author}{\bibfnamefont{S.~P.} \bibnamefont{{Gerhardt}}},
  \bibnamefont{and} \bibinfo{author}{\bibfnamefont{G.~J.}
  \bibnamefont{{Kramer}}}, \bibinfo{journal}{Physics of Plasmas}
  \textbf{\bibinfo{volume}{20}}, \bibinfo{pages}{082502}
  (\bibinfo{year}{2013}).

\bibitem[{\citenamefont{{Farmer} and {Morales}}(2013)}]{2013PhPl...20h2132F}
\bibinfo{author}{\bibfnamefont{W.~A.} \bibnamefont{{Farmer}}} \bibnamefont{and}
  \bibinfo{author}{\bibfnamefont{G.~J.} \bibnamefont{{Morales}}},
  \bibinfo{journal}{Physics of Plasmas} \textbf{\bibinfo{volume}{20}},
  \bibinfo{pages}{082132} (\bibinfo{year}{2013}).

\bibitem[{\citenamefont{{Heyvaerts} and {Priest}}(1983)}]{1983A&A...117..220H}
\bibinfo{author}{\bibfnamefont{J.}~\bibnamefont{{Heyvaerts}}} \bibnamefont{and}
  \bibinfo{author}{\bibfnamefont{E.~R.} \bibnamefont{{Priest}}},
  \bibinfo{journal}{Astron. Astrophys.} \textbf{\bibinfo{volume}{117}},
  \bibinfo{pages}{220} (\bibinfo{year}{1983}).

\bibitem[{\citenamefont{{Hood} et~al.}(2002)\citenamefont{{Hood}, {Brooks}, and
  {Wright}}}]{2002RSPSA.458.2307W}
\bibinfo{author}{\bibfnamefont{A.~W.} \bibnamefont{{Hood}}},
  \bibinfo{author}{\bibfnamefont{S.~J.} \bibnamefont{{Brooks}}},
  \bibnamefont{and} \bibinfo{author}{\bibfnamefont{A.~N.}
  \bibnamefont{{Wright}}}, \bibinfo{journal}{Royal Society of London
  Proceedings Series A} \textbf{\bibinfo{volume}{458}}, \bibinfo{pages}{2307}
  (\bibinfo{year}{2002}).

\bibitem[{\citenamefont{{Tsiklauri} et~al.}(2003)\citenamefont{{Tsiklauri},
  {Nakariakov}, and {Rowlands}}}]{2003A&A...400.1051T}
\bibinfo{author}{\bibfnamefont{D.}~\bibnamefont{{Tsiklauri}}},
  \bibinfo{author}{\bibfnamefont{V.~M.} \bibnamefont{{Nakariakov}}},
  \bibnamefont{and}
  \bibinfo{author}{\bibfnamefont{G.}~\bibnamefont{{Rowlands}}},
  \bibinfo{journal}{Astron. Astrophys.} \textbf{\bibinfo{volume}{400}},
  \bibinfo{pages}{1051} (\bibinfo{year}{2003}).

\bibitem[{\citenamefont{{Similon} and {Sudan}}(1989)}]{1989ApJ...336..442S}
\bibinfo{author}{\bibfnamefont{P.~L.} \bibnamefont{{Similon}}}
  \bibnamefont{and} \bibinfo{author}{\bibfnamefont{R.~N.}
  \bibnamefont{{Sudan}}}, \bibinfo{journal}{\apj}
  \textbf{\bibinfo{volume}{336}}, \bibinfo{pages}{442} (\bibinfo{year}{1989}).

\bibitem[{\citenamefont{{De Moortel} et~al.}(2000)\citenamefont{{De Moortel},
  {Hood}, and {Arber}}}]{2000A&A...354..334D}
\bibinfo{author}{\bibfnamefont{I.}~\bibnamefont{{De Moortel}}},
  \bibinfo{author}{\bibfnamefont{A.~W.} \bibnamefont{{Hood}}},
  \bibnamefont{and} \bibinfo{author}{\bibfnamefont{T.~D.}
  \bibnamefont{{Arber}}}, \bibinfo{journal}{Astron. Astrophys.}
  \textbf{\bibinfo{volume}{354}}, \bibinfo{pages}{334} (\bibinfo{year}{2000}).

\bibitem[{\citenamefont{{Smith} et~al.}(2007)\citenamefont{{Smith},
  {Tsiklauri}, and {Ruderman}}}]{2007A&A...475.1111S}
\bibinfo{author}{\bibfnamefont{P.~D.} \bibnamefont{{Smith}}},
  \bibinfo{author}{\bibfnamefont{D.}~\bibnamefont{{Tsiklauri}}},
  \bibnamefont{and} \bibinfo{author}{\bibfnamefont{M.~S.}
  \bibnamefont{{Ruderman}}}, \bibinfo{journal}{Astron. Astrophys.}
  \textbf{\bibinfo{volume}{475}}, \bibinfo{pages}{1111} (\bibinfo{year}{2007}).

\bibitem[{\citenamefont{{Malara} et~al.}(2000)\citenamefont{{Malara},
  {Petkaki}, and {Veltri}}}]{2000ApJ...533..523M}
\bibinfo{author}{\bibfnamefont{F.}~\bibnamefont{{Malara}}},
  \bibinfo{author}{\bibfnamefont{P.}~\bibnamefont{{Petkaki}}},
  \bibnamefont{and} \bibinfo{author}{\bibfnamefont{P.}~\bibnamefont{{Veltri}}},
  \bibinfo{journal}{\apj} \textbf{\bibinfo{volume}{533}}, \bibinfo{pages}{523}
  (\bibinfo{year}{2000}).

\bibitem[{\citenamefont{{Boozer}}(2012{\natexlab{a}})}]{2012PhPl...19k2901B}
\bibinfo{author}{\bibfnamefont{A.~H.} \bibnamefont{{Boozer}}},
  \bibinfo{journal}{Phys. Plasmas} \textbf{\bibinfo{volume}{19}},
  \bibinfo{pages}{112901} (\bibinfo{year}{2012}{\natexlab{a}}).

\bibitem[{\citenamefont{{Boozer}}(2012{\natexlab{b}})}]{2012PhPl...19i2902B}
\bibinfo{author}{\bibfnamefont{A.~H.} \bibnamefont{{Boozer}}},
  \bibinfo{journal}{Phys. Plasmas} \textbf{\bibinfo{volume}{19}},
  \bibinfo{pages}{092902} (\bibinfo{year}{2012}{\natexlab{b}}).

\bibitem[{\citenamefont{{Grappin} et~al.}(2008)\citenamefont{{Grappin},
  {Aulanier}, and {Pinto}}}]{2008A&A...490..353G}
\bibinfo{author}{\bibfnamefont{R.}~\bibnamefont{{Grappin}}},
  \bibinfo{author}{\bibfnamefont{G.}~\bibnamefont{{Aulanier}}},
  \bibnamefont{and} \bibinfo{author}{\bibfnamefont{R.}~\bibnamefont{{Pinto}}},
  \bibinfo{journal}{Astron. Astrophys.} \textbf{\bibinfo{volume}{490}},
  \bibinfo{pages}{353} (\bibinfo{year}{2008}).

\bibitem[{\citenamefont{{Arber} et~al.}(2001)\citenamefont{{Arber},
  {Longbottom}, {Gerrard}, and {Milne}}}]{2001JCoPh.171..151A}
\bibinfo{author}{\bibfnamefont{T.~D.} \bibnamefont{{Arber}}},
  \bibinfo{author}{\bibfnamefont{A.~W.} \bibnamefont{{Longbottom}}},
  \bibinfo{author}{\bibfnamefont{C.~L.} \bibnamefont{{Gerrard}}},
  \bibnamefont{and} \bibinfo{author}{\bibfnamefont{A.~M.}
  \bibnamefont{{Milne}}}, \bibinfo{journal}{Journal of Computational Physics}
  \textbf{\bibinfo{volume}{171}}, \bibinfo{pages}{151} (\bibinfo{year}{2001}).

\bibitem[{mov()}]{mov}
\emph{\bibinfo{title}{See supplemental material: 
 [Movie 1 \url{http://astro.qmul.ac.uk/~tsiklauri/abc_mov1.wmv}
  that shows time dynamics of $B_y(x,y=y_{max}/2,z)$
  contour plot for the case of AW Gaussian pulse launched along uniform
  background magnetic field. Movie 2 \url{http://astro.qmul.ac.uk/~tsiklauri/abc_mov2.wmv}
  that shows time dynamics of
  $B_y(x,y=y_{max}/2,z)$ contour plot for the case of harmonic AW launched
  along uniform background magnetic field. In both Movies 1 and 2 horizontal
  axis is the $x$-coordinate and vertical axis is $z$-coordinate. 
  Movie 3 \url{http://astro.qmul.ac.uk/~tsiklauri/abc_mov3.wmv} that
  shows time dynamics of $B_y(x,y=y_{max}/2,z,t)-B_{y0}(x,y=y_{max}/2,z,t)$
  shaded surface plot, i.e. difference between the magnetic field y-component in
  the case of ABC field without an AW but with resistivity $\hat
  \eta=5\times10^{-4}$ (here denoted by $B_y(x,y=y_{max}/2,z)$) and magnetic field
  y-component in the case of ABC field without an AW pulse and without
  resistivity $\hat \eta=0$ (here denoted by $B_{y0}(x,y=y_{max}/2,z)$). 
  Movie 4 \url{http://astro.qmul.ac.uk/~tsiklauri/abc_mov4.wmv}
  time dynamics of $B_y(x,y=y_{max}/2,z,t)-B_{y0}(x,y=y_{max}/2,z,t)$ shaded
  surface plot, i.e. difference between the magnetic field y-component in the
  case of ABC field with an AW pulse and with resistivity $\hat
  \eta=5\times10^{-4}$ (here denoted by $B_y(x,y=y_{max}/2,z)$) and magnetic field
  y-component in the case of ABC field without an AW pulse and with resistivity
  $\hat \eta=5\times10^{-4}$ (here denoted by $B_{y0}(x,y=y_{max}/2,z)$). 
  Movie 5 \url{http://astro.qmul.ac.uk/~tsiklauri/abc_mov5.wmv}
  is the same as Movie 4, but  with the harmonic AW instead of the Gaussian pulse.]}}.

\bibitem[{\citenamefont{{Aarts} and {Ooms}}(1998)}]{1998JEnMa..34..435A}
\bibinfo{author}{\bibfnamefont{A.~C.~T.} \bibnamefont{{Aarts}}}
  \bibnamefont{and} \bibinfo{author}{\bibfnamefont{G.}~\bibnamefont{{Ooms}}},
  \bibinfo{journal}{J. Eng. Math.}
  \textbf{\bibinfo{volume}{34}}, \bibinfo{pages}{435} (\bibinfo{year}{1998}).

\bibitem[{\citenamefont{{Kraichnan}}(1967)}]{1967PhFl...10.1417K}
\bibinfo{author}{\bibfnamefont{R.~H.} \bibnamefont{{Kraichnan}}},
  \bibinfo{journal}{Phys. Fluids} \textbf{\bibinfo{volume}{10}},
  \bibinfo{pages}{1417} (\bibinfo{year}{1967}).

\bibitem[{\citenamefont{{Sommeria}}(1986)}]{1986JFM...170..139S}
\bibinfo{author}{\bibfnamefont{J.}~\bibnamefont{{Sommeria}}},
  \bibinfo{journal}{J. Fluid Mech.} \textbf{\bibinfo{volume}{170}},
  \bibinfo{pages}{139} (\bibinfo{year}{1986}).

\bibitem[{\citenamefont{{Bard{\'o}czi}
  et~al.}(2012)\citenamefont{{Bard{\'o}czi}, {Berta}, and
  {Bencze}}}]{2012PhRvE..85e6315B}
\bibinfo{author}{\bibfnamefont{L.}~\bibnamefont{{Bard{\'o}czi}}},
  \bibinfo{author}{\bibfnamefont{M.}~\bibnamefont{{Berta}}}, \bibnamefont{and}
  \bibinfo{author}{\bibfnamefont{A.}~\bibnamefont{{Bencze}}},
  \bibinfo{journal}{Phys. Rev. E} \textbf{\bibinfo{volume}{85}},
  \bibinfo{eid}{056315} (\bibinfo{year}{2012}).


\end{thebibliography}

\end{document}